\documentclass[letterpaper,9pt]{article}
\usepackage[numbers]{natbib} % Use natbib with numbers for citations
\usepackage{tcolorbox}
\tcbuselibrary{breakable}
\usepackage{authblk}
\usepackage{caption}
\usepackage{tabularx} % extra features for tabular environment
\usepackage{amsmath,amssymb,slashed}   % improve math presentation
\usepackage{graphicx} % takes care of graphic including machinery
\usepackage[margin=1in,letterpaper]{geometry} % decreases margins
\usepackage{tcolorbox} % for creating the boxed environment
\usepackage{xcolor} % for coloring the text
\usepackage{hyperref} 
\usepackage{dirtytalk}
\usepackage{enumitem}
\usepackage{csquotes}

\tcbuselibrary{breakable}
\usepackage{pythonhighlight}
\usepackage{cancel}
\usepackage{multirow}
\usepackage{booktabs}
\usepackage{listings}

\usepackage{pgfplots}
\usepackage{tikz}
\usepackage{multirow}
\usepackage{booktabs}
\usepackage{bm}

\usepackage{listings}

\newtcolorbox{llmreport}{
  colback=gray!5,
  colframe=gray!60,
  title=LLM-Generated Algorithmic Summary,
  fonttitle=\bfseries,
  breakable
}
\title{MadEvolve: Evolutionary Optimization of Cosmological Algorithms with Large Language Models}

\author[1]{Tianyi Li} 
\author[1]{Shihui Zang} 
\author[1]{Moritz M\"unchmeyer}

\affil[1]{Department of Physics, University of Wisconsin-Madison}

\begin{document} 
\maketitle
    
\begin{abstract}

We develop a general framework to discover scientific algorithms and apply it to three problems in computational cosmology. Our code, \emph{MadEvolve}, is similar to Google's AlphaEvolve, but places a stronger emphasis on free parameters and their optimization. Our code starts with a baseline human algorithm implementation, and then optimizes its performance metrics by making iterative changes to its code. As a further convenient feature, MadEvolve automatically generates a report that compares the input algorithm with the evolved algorithm, describes the algorithmic innovations and lists the free parameters and their function. Our code supports both auto-differentiable, gradient-based parameter optimization and gradient-free optimization methods. We apply MadEvolve to the reconstruction of cosmological initial conditions, 21cm foreground contamination reconstruction and effective baryonic physics in N-body simulations. In all cases, we find substantial improvements over the base algorithm. We make MadEvolve and our three tasks publicly available at \url{madevolve.org}.

\end{abstract}

\vspace{10cm}
\begin{center}
\normalsize
    Corresponding Author: Tianyi Li (\texttt{tianyi.li@wisc.edu})
\end{center}

\clearpage

\setcounter{tocdepth}{2}
\tableofcontents
\clearpage

\section{Introduction}

Modern cosmological data analysis involves numerous algorithmic challenges, such as field reconstruction, correlation function estimation, simulations, and many others across different tracers and their combinations. While human researchers continue to make algorithmic advances on these tasks, traditional approaches based on handcrafted algorithms and manual hyperparameter tuning struggle with the growing data volumes, model complexity, and exponentially many design choices in large pipelines. We need systematic methods for automated algorithm discovery that can search through conceptual spaces while respecting physical constraints and computational limits.

Our work primarily builds on two recent breakthroughs combining large language models (LLMs) with evolutionary search. \emph{FunSearch} \cite{RomeraParedes2024FunSearch} demonstrated that pairing a code-trained LLM with an automated evaluator could produce genuinely novel results---including new cap-set constructions and better bin-packing heuristics---by searching through function space using best-shot prompting and island-based populations for diversity. \emph{AlphaEvolve} \cite{AlphaEvolve2025} extended this approach to create an evolutionary coding agent capable of modifying entire codebases through LLM ensembles and automated metrics, achieving improvements on both mathematical problems and real-world system optimizations. 

We adapt these ideas to cosmology, treating reconstruction and analysis programs themselves as the optimization target. LLMs serve as smart mutation operators that understand code semantics, while physics-based evaluators guide the search toward domain-specific objectives. Our main contributions are: 

\begin{enumerate}[label=(\roman*)]
\item We develop the \emph{MadEvolve} framework, integrating LLM-driven evolution with physical fitness metrics and HPC-compatible evaluation. MadEvolve is publicly available at \url{madevolve.org}. We believe that MadEvolve, as a light-weight open-source framework with extensive features, will be a useful tool beyond cosmology.
\item MadEvolve includes a \emph{nested optimization architecture} where an inner loop automatically tunes continuous parameters (e.g. filter scales and expansion coefficients) via gradient-based or grid-search methods whenever the LLM proposes new algorithmic structures, separating structural search from parameter optimization. 
\item MadEvolve automatically generates a \emph{scientific report} by using LLMs to analyze discovered algorithms, compare them against the baseline, and produce human-readable explanations of the physical intuition behind improvements.
\item We develop \emph{three applications} to common tasks in cosmological data analysis: initial condition reconstruction (BAO reconstruction), 21 cm foreground removal reconstruction (i.e. recovering modes lost due to foregrounds), and effective baryonic physics in N-body simulations (at the example of thermal Sunyaev-Zeldovich effect prediction), showing that the system independently discovers novel algorithmic strategies. In all cases, the performance metric is evaluated at each evolution step on a small set of training and validation simulations. As we will show, our resulting  algorithms improve over the baseline algorithm in all three cases. This includes setups where the human baseline is strong (e.g. iterative initial condition reconstruction). A summary of these achievements can be found in Sec. \ref{sec:overview_of_results}. Our three optimization tasks, which were designed to be computationally tractable for thousands of calls, are also made available at \url{madevolve.org}.
\end{enumerate}

%Our results show that the presented approach to solving problems in computational  cosmology is very powerful. 
Developing algorithms is a complementary and in many cases superior approach to black-box machine learning (which often uses millions or billions of parameters). In cosmology, supervised machine learning on entire simulation boxes is difficult to deploy in practice because it is hard to simulate large-scale surveys with sufficient accuracy and volume. On the other hand, our algorithms have only a \emph{small number of parameters and generalize well} outside of the training setup (which we test in a limit setup using different simulation resolutions). Further, these algorithms are humanly interpretable, which can be useful to gauge their reliability. Because they are interpretable, they can also be used as \emph{verified idea generators} for human scientists: Once a strong algorithm is found, a human researcher can inspect how it works and possibly build on it.

The presented use cases for LLM reasoning circumvent the problem of the lack of reliability of LLMs when solving theoretical problems. As was shown e.g. in \cite{Chung:2025nsd,Gao:2025tdy}, LLMs face significant challenges in generating error-free derivations or calculations in theoretical physics. More generally, even frontier LLM models struggle with consistent reasoning performance in both formal and informal reasoning, with significant failure modes even for some seemingly simple tasks (see \cite{2026arXiv260206176S} and references therein). However, the tasks in this work are \emph{verifiable}, i.e. success can be measured by the performance metric achieved by the algorithm on novel test data. Verifiability is in general difficult to achieve in theoretical physics. While parts of mathematics have been formalized (e.g. with the LEAN language \cite{10.1007/978-3-030-79876-5_37}), and new LLM-based proofs can thus sometimes be auto-verified, this is much harder to achieve in theoretical physics. Fortunately, in the present case, the validity of new LLM-generated algorithmic ideas can be checked on simulations.

While our tasks can in principle also be solved with agentic frameworks (see e.g. the Denario project for agent applications to cosmology \cite{Denario2025}), the combination of LLMs with evolutionary techniques allows to systematically explore independent ideas, and ultimately combine them into a best-performing algorithm (rather than risk getting stuck in a local minimum). There are many promising directions to further improve such systems, as we discuss below in Sec. \ref{sec:future_directions}. 

The paper is organized as follows: In Section~\ref{sec:related}, we review related work on LLM-guided code evolution and machine learning in cosmology. Section~\ref{sec:framework} introduces the MadEvolve framework and its core components. Section~\ref{sec:overview_of_results} provides an overview of our main results across all three applications. Sections~\ref{sec:app1}, \ref{sec:app2}, and \ref{sec:app3} present detailed analyses of each application: initial condition and BAO reconstruction, 21cm foreground contamination reconstruction, and effective baryonic physics in N-body simulations, respectively. Section~\ref{sec:future_directions} discusses future directions to improve MadEvolve, and Section~\ref{sec:conclusion} concludes. Appendix~\ref{app:implementation_details} provides implementation details, Appendix~\ref{app:new_tasks} describes how to apply MadEvolve to new tasks, and Appendix~\ref{app:llm_report} refers to LLM-generated evolution reports for all tasks.

\section{Related Work}
\label{sec:related}

\subsection{Program Synthesis and Repair with LLMs and Agents}
Recent frontier LLMs increasingly support end-to-end software development workflows: generating and refactoring code from natural language, debugging with long-context understanding, and using tools to iterate across multi-step projects. OpenAI's GPT-5 series emphasizes long-horizon reasoning and agentic tool-calling, with strong reported gains on software-engineering style evaluations and practical coding workflows \cite{OpenAI_GPT52}. Complementing this, reasoning-oriented model families such as OpenAI o1 and o3 are explicitly designed to ``think longer'' for difficult tasks, improving reliability on coding- and STEM-heavy problems \cite{OpenAI_o1,OpenAI_o3_o4mini}. Google's Gemini 3 represents a major step toward AGI, combining state-of-the-art reasoning with advanced multimodal understanding and agentic functionality \cite{Gemini3_2025}. Anthropic's Claude Opus 4.5 achieves industry-leading results on coding benchmarks such as SWE-bench and is the best-performing model for computer use and agentic tasks \cite{Claude45_2025}.

Beyond synthesis, \emph{code repair} has become a central benchmark and application domain. SWE-bench evaluates systems on 2,294 real-world GitHub issues across 12 popular Python repositories, requiring coordinated edits across files and execution of test suites \cite{Jimenez2024SWEbench,SWEBenchSite}. This setting has catalyzed agentic coding systems that plan, edit repositories, and run tools in an iterative loop. SWE-agent introduces an agent--computer interface that enables language models to navigate repositories, modify files, and execute tests more effectively \cite{Yang2024SWEagent}. In industry-facing tooling, Anthropic's Claude Code integrates directly into terminals and IDEs and adds features such as checkpoints and more autonomous workflows for complex development tasks \cite{AnthropicClaudeCode2025}. Systems like Devin \cite{CognitionDevin2024} further illustrate end-to-end autonomous software engineering in realistic repo environments, highlighting both the promise and remaining reliability challenges of fully autonomous agents.

\subsection{Evolutionary and Quality–Diversity (QD) Search}

Following AlphaEvolve \cite{AlphaEvolve2025}, our work draws heavily from evolutionary algorithms and quality-diversity (QD) methods in machine learning. QD algorithms like \emph{MAP-Elites} \cite{Mouret2015, Cully2015, Pugh2015} maintain a diverse archive of good solutions organized by behavioral traits, capturing different strategies and trade-offs. This approach has proven effective in robotics and game AI, and has become a core component of modern LLM-guided evolution systems including AlphaEvolve \cite{AlphaEvolve2025}. For algorithm discovery, QD helps find unexpected solutions that single-objective optimization might miss by explicitly maintaining population diversity.

Recent work has combined evolution with domain knowledge to discover new algorithms. AutoML-Zero evolved ML algorithms from scratch, rediscovering gradient descent and other fundamentals \cite{Real2020}. AlphaTensor and AlphaDev showed that search guided by learned models can beat human-designed algorithms for matrix multiplication and sorting \cite{Fawzi2022, Mankowitz2023}. The key insight enabling LLM-guided evolution is using language models as intelligent mutation operators. FunSearch \cite{RomeraParedes2024FunSearch} pioneered this approach, discovering new solutions to the cap set problem---perhaps the first time an LLM contributed to solving an open mathematical problem. AlphaEvolve \cite{AlphaEvolve2025} extended this paradigm to evolve entire codebases rather than single functions. \textsc{Eureka} \cite{Lee2023Eureka} applied similar ideas to reward function design for robotics. More recently, EvoLLM \cite{EvoLLM2024} demonstrated that LLMs can serve as in-context recombination operators for black-box optimization, while LLaMEA \cite{LLaMEA2024} automates metaheuristic algorithm generation through iterative LLM refinement. These systems collectively establish that LLMs can act as context-aware mutators, combining broad exploration with informed modifications guided by execution feedback.

\subsection{Automated Discovery in the Physical Sciences}

Scientists have long sought to automate the discovery of physical laws and optimal methods from data. \emph{Symbolic regression} tries to find analytical formulas that explain datasets. Traditional methods of symbolic regression are often based on genetic programming. Symbolic regression has for example been used to learn conservation laws from simulations \cite{Cranmer2020SR}. Specifically in physics, AI Feynman \cite{Udrescu2020} rediscovered physics equations by combining neural networks with physics-inspired simplifications. 

More recently, several groups and companies have developed LLM-based \enquote{AI-Scientists}. These systems generate hypotheses, test them, and write publication-style summaries of their results, typically using multiple LLM-agents with different responsibilities. Sakana AI's \emph{The AI Scientist} \cite{AIScientist2024} was the first comprehensive framework for fully automated scientific discovery, generating research ideas, writing code, executing experiments, and producing complete papers at a cost of \$6--15 per paper. Its successor, \emph{The AI Scientist-v2} \cite{AIScientistV2_2025}, uses agentic tree search to produce the first entirely AI-generated peer-reviewed workshop paper. Google's \emph{AI Co-Scientist} \cite{GoogleCoScientist2025} is a multi-agent system built on Gemini 2.0 that accelerates biomedical discovery, with wet-lab validated predictions for drug repurposing in acute myeloid leukemia and epigenetic targets for liver fibrosis. The \emph{Denario} project \cite{Denario2025} focuses on astrophysics and cosmology, demonstrating AI agents that can perform literature review, code development, and paper drafting across multiple scientific disciplines. Other notable systems include \emph{Agent Laboratory} \cite{AgentLaboratory2025}, which uses specialized agents (PhD, Postdoc, Professor) to assist human researchers in executing their ideas, and \emph{Data-to-Paper} \cite{DataToPaper2024}, which creates verifiable manuscripts from annotated data with information tracing for transparency. 

Despite this impressive progress, AI-Scientist results remain limited. This is likely because LLMs struggle to judge both the importance and the correctness of their own results. In our work, we therefore apply LLMs to human-defined tasks with a clear verifiable reward metric, and let the LLM reason within these more limiting constraints. 

\section{The MadEvolve Framework}
\label{sec:framework}

\subsection{Overview}

Our evolutionary framework treats scientific algorithms as programs which are iteratively improved by language models. The core loop samples a parent program from a diverse population, prompts an LLM for modifications, evaluates against physics-based metrics, and updates the population. We maintain diversity through MAP-Elites grids and island-based subpopulations, while LLM ensembles generate varied code changes, from targeted \enquote{diff patches} to complete rewrites.

Our design is based on AlphaEvolve \cite{AlphaEvolve2025}, which demonstrated breakthrough discoveries in mathematics and engineering by combining Gemini models with evolutionary search. AlphaEvolve employs sophisticated infrastructure: asynchronous distributed evaluation, meta prompt evolution, and multi-score optimization. Several open-source implementations have since emerged. ShinkaEvolve \cite{lange2025shinkaevolve} emphasizes sample efficiency through novelty rejection sampling with cosine similarity filtering, UCB1-style bandit algorithms for dynamic LLM selection, and a meta-scratchpad distilling insights from successful mutations. OpenEvolve~\cite{OpenEvolve2025} provides reproducibility features including hash-based run isolation, multi-language support, and flexible API integration.

MadEvolve is similar but slightly simplified compared to AlphaEvolve and ShinkaEvolve. We omit meta prompt evolution and explicit novelty filtering in favor of letting physics-based metrics guide the search directly. However, two distinctive features differentiate our framework. First, \emph{budget-constrained inner optimization loops} automatically tune continuous parameters before computing fitness, ensuring fair comparisons between algorithms at their best achievable performance. The system offers two complementary strategies: greedy one-shot tuning via derivative-free search for general programs with non-differentiable operations, and autodiff-based optimization using Adam for fully differentiable pipelines where gradient information enables efficient navigation of high-dimensional parameter spaces. Second, \emph{automated report generation} produces human-readable analyses through a three-stage pipeline: \emph{lineage extraction} traces the evolutionary ancestry of discovered algorithms, \emph{comparative analysis} invokes an LLM to identify structural differences and hypothesize physical rationale, and \emph{synthesis} aggregates these into coherent narratives with annotated code comparisons and performance visualizations.

% -------------------- framework figure --------------------
\begin{figure}[htbp]
    \centering
    \includegraphics[width=0.58\linewidth]{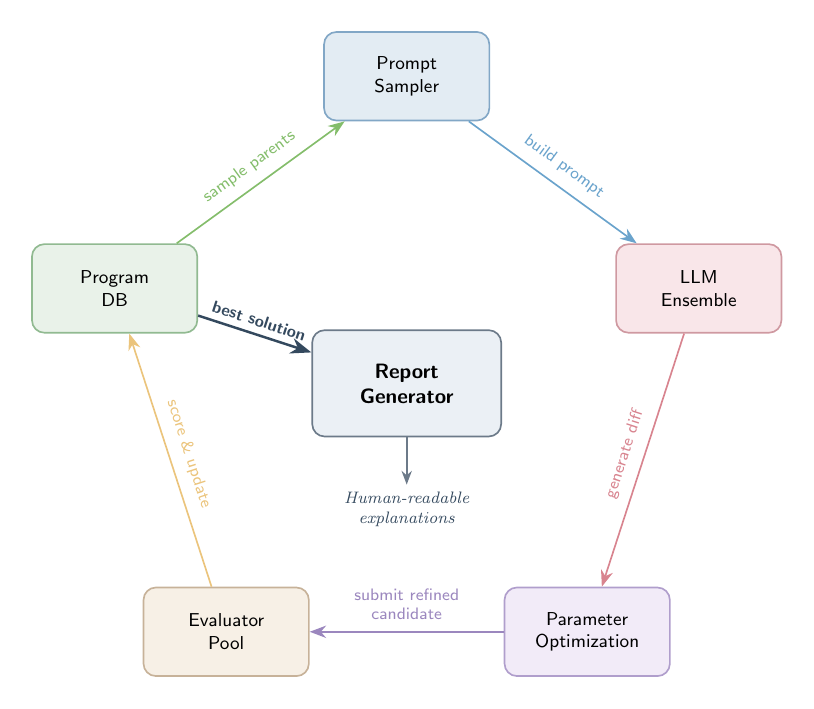}
    \caption{Overview of the LLM-driven evolutionary pipeline in MadEvolve.
    The cycle proceeds as a closed loop: the \emph{Prompt Sampler} retrieves parent programs from the \emph{Program DB} to query the \emph{LLM Ensemble}.
    The generated code diffs first undergo \emph{Parameter Optimization} to refine continuous variables before being submitted to the \emph{Evaluator Pool}.
    Evaluation scores update the database, completing the iteration.
    Finally, the best solution is transmitted to the central \emph{Report Generator} to produce human-readable explanations of the discovered innovations.}
    \label{fig:alphaevolve_flowchart}
\end{figure}

\subsection{Core Architecture and Evolution Loop}

The central controller orchestrates the evolutionary process through an asynchronous loop coordinating sampling, generation, evaluation, and selection. Each iteration proceeds through six stages. 

\begin{itemize}

\item \emph{Parent selection} draws a program from the population, balancing exploitation of high-fitness solutions against exploration through inter-island migration (Section~\ref{subsec:population}).

\item \emph{Inspiration retrieval} gathers $k$ exemplar programs including the global best, recent top performers, and structurally diverse neighbors from the MAP-Elites grid (Section~\ref{subsec:population}).

\item In \emph{prompt construction}, the system assembles a prompt containing the parent's source code, performance metrics, identified areas for improvement, and retrieved exemplars. \emph{Variation generation} then queries the LLM ensemble for code modifications---either targeted diff patches for incremental refinement or complete rewrites for exploring different approaches (Section~\ref{subsec:llm_ensemble}).

\item The candidate undergoes \emph{fitness evaluation} against the task-specific protocol. When the program contains tunable parameters (e.g., smoothing scales, learning rates), an inner \emph{optimization loop} automatically calibrates these values before computing fitness (Section~\ref{subsec:inner_opt}).

\item Finally, \emph{population update} incorporates the program into the database, updating the MAP-Elites grid, island populations, and elite archive (Section~\ref{subsec:population}). The loop continues until reaching a target fitness, exhausting the iteration budget, or detecting stagnation.

\end{itemize}

\subsection{Parameter Tracking and Optimization}
\label{subsec:inner_opt}

Compared to the original AlphaEvolve, our work 
emphasizes the role of tunable parameters. We aim to find algorithms that have as few as possible free parameters (which usually improves generalization), and we track their meaning and optimize their numerical values. When the LLM proposes a new algorithm, it often introduces tunable parameters (smoothing scales, threshold values, iteration counts) whose optimal values are not known in advance. If we evaluate the algorithm with poorly chosen parameters, we might reject a genuinely good idea simply because it was not properly tuned. On the other hand, running extensive optimization for every candidate would be prohibitively expensive. We resolve this tension through a \emph{budget-constrained inner optimization loop} that tunes parameters automatically while keeping computational costs under control. Depending on the nature of the algorithm and its parameters, we offer two complementary optimization strategies: greedy one-shot tuning for general programs and autodiff-based optimization for fully differentiable pipelines.

\subsubsection{Greedy One-Shot Optimization}

For general algorithms that may contain non-differentiable operations, we employ a \emph{greedy one-shot} protocol: each newly introduced parameter gets tuned exactly once when it first appears, after which its value is frozen. The system detects new parameters by comparing the child program's parameter set against its parent. Only parameters that appear for the first time trigger optimization; inherited parameters retain their previously optimized values. 

To enable automatic optimization, evolved programs declare their tunable parameters in a configuration block, specifying for each parameter: an initial value, the optimization method (e.g., \texttt{grid} for derivative-free search), and valid bounds. We perform derivative-free search within a strict budget of $B$ evaluations, where the default budget and search strategy are specified in the system configuration file. For instance, with a budget of 10 evaluations, one or two parameters can be tuned via grid search, while larger parameter sets may require Latin hypercube sampling or Bayesian optimization with a Gaussian process surrogate to focus evaluations on promising regions.

After optimization, the best values found are injected back into the program source code and the parameters become frozen for subsequent generations. This protocol keeps optimization costs low---not every mutation introduces new parameters, and when optimization does occur, the overhead typically amounts to only a small multiple of a single evaluation cost. If the LLM later decides that a frozen parameter needs adjustment, perhaps because subsequent structural changes have shifted the optimal value, it must remove and reintroduce the parameter through a code rewrite. This triggers a fresh optimization cycle. The mechanism prevents parameters from drifting indefinitely while still allowing the system to revisit tuning decisions when the algorithm structure changes significantly.

While we used one-shot tuning in early stages of this work, the final applications described below all make use of fully auto-differentiable algorithms, as described in the next section.

\subsubsection{Auto-differentiable Parameter Optimization}

For algorithms where the target metric can be made differentiable with respect to the algorithm parameters, we offer an alternative optimization strategy based on automatic differentiation that is substantially more powerful for continuous parameters. This approach requires translating the initial program into a fully differentiable form (e.g., using JAX), after which gradients of the fitness function with respect to all continuous parameters can be computed efficiently through reverse-mode automatic differentiation. As in the one-shot case, each program declares its tunable parameters in a configuration block, now specifying \texttt{autodiff} as the optimization method along with initial values and valid bounds.

Before each fitness evaluation, the system runs gradient-based optimization on all parameters marked for autodiff tuning. Using the Adam optimizer, the system iteratively adjusts parameter values to maximize fitness, typically converging within 10--50 iterations. Once optimized, the parameter values are injected back into the program source for subsequent evaluations.

The autodiff approach offers key advantages over derivative-free methods: gradient information enables efficient navigation of high-dimensional parameter spaces, allowing simultaneous optimization of 10+ continuous parameters within dozens of iterations rather than exponentially many function evaluations. The main requirement is that the evolved algorithm must use differentiable operations throughout. While in principle this includes the case of training a large neural network, MadEvolve tracks individual parameters and is not meant for neural network training (as we indicate to the LLMs in their prompts).

\subsubsection{Choosing Between Strategies }

The choice between greedy one-shot tuning and autodiff optimization depends on the problem structure. Autodiff is preferred when: (i) the fitness metric is smooth and differentiable with respect to parameters, (ii) the algorithm involves primarily linear algebra, FFTs, and smooth nonlinearities, and (iii) multiple continuous parameters need simultaneous optimization. Greedy one-shot tuning is preferred when: (i) the algorithm contains fundamentally discrete choices or non-differentiable operations and the number of tunable parameters is small, or (ii) rapid prototyping is prioritized over optimal parameter values. Both strategies share the same goal---ensuring that fitness comparisons are fair by evaluating each algorithm at its best achievable performance rather than penalizing suboptimal initial parameter choices. MadEvolve supports mixed algorithms, where some parameters are auto-differentiable and others are not.

\subsection{Hybrid Population Management}
\label{subsec:population}

Population management is a core component of LLM-guided evolutionary systems and has been addressed in different ways by prior work. AlphaEvolve~\cite{AlphaEvolve2025} combines a MAP-Elites quality-diversity archive~\cite{Mouret2015} with an island model, using three fixed feature dimensions (performance, code diversity via Levenshtein distance, and complexity via code length) and a default configuration of 5 islands with up to 25 programs each. OpenEvolve~\cite{OpenEvolve2025} generalizes this with configurable feature dimensions, dynamic feature scaling, and adaptive bin counts, while maintaining the same island-based architecture with ring-topology migration. ShinkaEvolve~\cite{lange2025shinkaevolve} takes a different approach, relying solely on an island model without MAP-Elites, and instead maintaining diversity through novelty-based rejection sampling that uses embedding similarity combined with an LLM-as-judge to filter near-duplicate proposals before evaluation. MadEvolve adopts a hybrid strategy that draws from all three systems but simplifies certain aspects that we found unnecessary for our physics applications. Our population database employs three complementary mechanisms.

The first is a \emph{MAP-Elites grid}~\cite{Mouret2015}, which partitions behavioral space into an $n$-dimensional structure $\mathcal{G} = \prod_{i=1}^{n} [0, b_i) \subset \mathbb{Z}^n$. Following the feature dimensions used by AlphaEvolve and OpenEvolve, we define three axes: \emph{complexity} (code character length), \emph{diversity} (mean cosine distance of the program's code embedding to a reservoir-sampled reference set of representative programs), and \emph{performance} (the combined fitness score). Each cell retains exactly one program; a new program replaces the incumbent only when demonstrating superior performance. We chose MAP-Elites over the novelty rejection sampling approach of ShinkaEvolve because it provides a persistent archive of structurally diverse solutions that can serve as inspiration during prompt construction, rather than only filtering at generation time. For our cosmological applications, where algorithmic strategies can differ qualitatively (e.g., single-pass versus multi-pass reconstruction, different physical correction terms), preserving this diversity explicitly in the archive proved more effective than relying on embedding-based duplicate detection.

The second mechanism is an \emph{island model}, which partitions the population into $m$ semi-isolated subpopulations $\mathcal{I}_1, \ldots, \mathcal{I}_m$, a standard approach used in all three reference systems. Islands evolve independently while periodic migration events (every $T_\text{mig}$ generations, transferring a fraction $\rho$ of each island's best members to its neighbor via a ring topology) enable cross-pollination, preventing premature convergence by maintaining competing lineages. By default we set $m=5$, $T_\text{mig}=5$, and $\rho=0.1$. We use a fixed-interval migration schedule as in AlphaEvolve, rather than OpenEvolve's event-driven lazy migration (triggered by program addition counts), as we found fixed intervals simpler to reason about and sufficient for our problem sizes. Unlike ShinkaEvolve, we do not enforce island elitism (protecting the best program on each island from being replaced by migrants), since in our experience the global elite archive (described below) already preserves top performers.

The third mechanism maintains a \emph{global elite archive} $\mathcal{A}$ of fixed capacity that preserves the highest-performing programs by fitness score; when the archive is full, a new program is admitted only if it surpasses the current worst entry. This component is not present in ShinkaEvolve, which relies on per-island archives, and serves as an additional safeguard against losing high-performing solutions during migration or MAP-Elites cell replacement. Together, these three mechanisms ensure both breadth (MAP-Elites diversity and island isolation) and depth (elite preservation and local exploitation).

Compared to AlphaEvolve, our setup omits meta-prompt evolution (where the system co-evolves the prompts themselves in a parallel database) and explicit novelty filtering. We found that for our physics applications, where the evaluator provides a strong and informative fitness signal, letting the physics-based metrics guide the search directly was sufficient to maintain productive evolution without the additional complexity of novelty detection.

\subsection{LLM Ensemble and Generation Strategies}
\label{subsec:llm_ensemble}

All three reference systems---AlphaEvolve~\cite{AlphaEvolve2025}, OpenEvolve~\cite{OpenEvolve2025}, and ShinkaEvolve~\cite{lange2025shinkaevolve}---employ ensembles of language models rather than relying on a single model, as different LLMs exhibit complementary strengths: fast, lightweight models (e.g., Gemini Flash) maximize throughput by testing many candidates, while larger, more capable models (e.g., GPT-5, Gemini Pro) occasionally propose higher-quality structural innovations. AlphaEvolve uses a fixed weighting of approximately 80\% Flash and 20\% Pro models. ShinkaEvolve goes further by treating model selection as a multi-armed bandit problem, dynamically routing queries to the model that has historically produced the largest fitness improvements, using a UCB1 strategy~\cite{auer2002finite} where the reward signal is defined as the relative improvement over the parent program's fitness. OpenEvolve uses simpler probabilistic model selection based on configured weights.

MadEvolve adopts the bandit-based approach of ShinkaEvolve, which we found to be more effective than fixed weights because the relative utility of different models changes over the course of evolution---lightweight models are often sufficient for early exploration, while later refinement of already-optimized code benefits from more capable models. Concretely, each model $M_i$ is treated as an arm in a UCB1 bandit~\cite{auer2002finite}. After every model has been queried at least once, the system computes
\begin{equation}
    \mathrm{UCB}_i \;=\; \Bigl(\bar{r}_i \;+\; c_t\,\sqrt{\frac{\ln N}{n_i}}\Bigr)\,w_i\,,
    \label{eq:ucb}
\end{equation}
where $\bar{r}_i$ is the empirical mean score improvement produced by model~$M_i$, $n_i$ the number of queries issued to it, $N = \sum_j n_j$ the total query count, and $w_i$ a prior base weight reflecting cost or capability preferences. The exploration coefficient decays geometrically as $c_t = c_0\,\gamma^{N}$ with initial value $c_0$ and decay rate $\gamma < 1$, so that the selector gradually shifts from exploration to exploitation as evidence accumulates. At each step the model with the highest $\mathrm{UCB}_i$ is selected greedily. Thompson Sampling and $\varepsilon$-greedy selection are available as alternative strategies. Unlike ShinkaEvolve, which defines the reward as $r_i = \exp(\max(r - r_{\text{base}}, 0)) - 1$ to incentivize bold improvements, we use the raw score improvement as the reward signal, which we found to be more stable for our physics applications where the fitness landscape is relatively smooth.

Regarding code generation modes, AlphaEvolve and OpenEvolve support both diff-based editing (using SEARCH/REPLACE blocks) and full code rewrites, with diff-based editing as the default mode. ShinkaEvolve adds a third \emph{crossover} mode in which the LLM receives code from multiple parent programs and performs guided recombination. MadEvolve's ensemble supports three analogous generation modes. In \emph{differential mode}, the LLM produces diff patches modifying specific code regions, which dominates during exploitation phases. In \emph{full rewrite mode}, the LLM generates complete programs from scratch, enabling radical departures from the current solution. In \emph{synthesis mode}, the LLM combines ideas from multiple reference programs to produce improved variants, similar to the crossover operator in ShinkaEvolve. By default, differential patches are selected with $\sim$70\% probability and full rewrites with $\sim$30\%, consistent with the exploitation/exploration ratios used in AlphaEvolve and OpenEvolve. The relative weights of all three modes are freely adjustable via configuration.

\subsection{Automated Report Generation}

A distinguishing capability of MadEvolve is automated interpretability: the system generates comprehensive reports analyzing discovered algorithms relative to baselines. Existing open-source evolutionary coding systems---including OpenEvolve~\cite{OpenEvolve2025} and ShinkaEvolve~\cite{lange2025shinkaevolve}---primarily output evolved code and numerical metrics (fitness scores, generation statistics), but do not automatically produce human-readable scientific analyses of \emph{what} the evolved algorithm does differently and \emph{why} those changes improve performance. OpenEvolve provides evolution tree visualization and per-generation statistics, and ShinkaEvolve offers an interactive web dashboard with genealogy trees and performance graphs, but neither system generates a narrative report explaining the algorithmic innovations in domain-specific terms. AlphaEvolve~\cite{AlphaEvolve2025} does not describe such a capability either. For scientific applications, where the goal is not merely to obtain a high-performing program but to \emph{understand} the principles behind its success, this interpretability gap is a significant limitation. A researcher receiving an evolved cosmological reconstruction algorithm needs to understand whether the improvements arise from known physical effects (e.g., second-order perturbation theory corrections) or genuinely novel strategies, and whether the approach generalizes beyond the training setup. MadEvolve addresses this gap through a three-stage automated report generation pipeline.

First, \emph{lineage extraction} traces the best program's full evolutionary ancestry back to the initial seed. Each program in the lineage is stored with its generation number, fitness score, the proposing LLM model name, and the diff that produced it (all recorded in the program metadata). The lineage reconstruction identifies the complete chain of mutations that transformed the baseline into the final algorithm, enabling attribution of specific innovations to particular evolutionary steps. The default report renders the evolutionary path with per-step score deltas, revealing whether progress was gradual (many small incremental improvements) or punctuated (a few large jumps interspersed with plateaus). This information is valuable for understanding the difficulty of the optimization landscape and for guiding future prompt engineering: if most progress came from a single breakthrough mutation, the prompt that triggered it can be studied and reused.

Second, \emph{comparative analysis} invokes an LLM to analyze the structural differences between the final and baseline algorithms. The system provides both complete code listings and a metrics comparison table to the LLM, which is prompted via a domain-specific adapter to identify key changes and hypothesize physical rationale for each modification, grounded in the application context. The domain-specific adapter is a configurable component that injects relevant scientific background into the analysis prompt---for example, for BAO reconstruction, it provides context about Lagrangian perturbation theory, displacement fields, and mode coupling, enabling the LLM to connect code-level changes (e.g., the introduction of deformation tensor invariants) to their physical interpretation (e.g., capturing second-order gravitational effects). This grounding in domain knowledge distinguishes the reports from generic code-diff summaries.

Third, \emph{synthesis} aggregates these analyses into a coherent narrative document including: a metrics comparison table with per-scale breakdowns, LLM-generated descriptions of the baseline and best algorithms, an improvement analysis highlighting what changed and why, an executive summary of the key findings, and annotated side-by-side code comparisons. The report is designed to serve as a starting point for human researchers, enabling rapid assessment of whether discovered algorithms merit further investigation, publication, or deployment in observational pipelines. In practice, we found that these reports substantially accelerated our own understanding of the evolved algorithms and reduced the time needed to identify which innovations were scientifically meaningful versus which may be artifacts of overfitting to the training metric. Nevertheless these reports do have some shortcomings inherited from the limited reasoning ability of current LLMs, as we comment on further below. More details of the generated reports and references to the reports for all three applications are provided in Appendix~\ref{app:llm_report}.

\section{Overview of Results}\label{sec:overview_of_results}

We applied MadEvolve to three distinct cosmological data analysis problems: baryon acoustic oscillation (BAO) reconstruction, 21\,cm foreground contamination mitigation via tidal field methods, and baryonic physics predictions from N-body dark matter simulations. For all these cases, we implemented base-algorithms from the literature and made them auto-differentiable with respect to their parameters. Across these applications, the system discovered algorithms that substantially outperform conventional baselines, with performance improvements ranging from 2.8\% to 30.7\% in cross-correlation metrics and up to 63\% reduction in prediction loss. Table~\ref{tab:overview_results} summarizes the key outcomes. The evolutionary runs required between 233 and 1{,}165 generations (see App. \ref{app:computation}), demonstrating that meaningful algorithmic improvements can be achieved without extensive infrastructure.

\begin{table}[h]
\centering
\begin{tabular}{lcccc}
\toprule
\textbf{Application} & \textbf{Baseline} & \textbf{Evolved} & \textbf{Improvement} & \textbf{Generations} \\
\midrule
BAO Reconstruction with StdRec ($\bar{r}_{\text{BAO}}$) & 0.752 & 0.924 & 22.8\% & 1{,}165 \\
BAO Reconstruction with IterRec ($\bar{r}_{\text{BAO}}$) & 0.933 & 0.959 & 2.8\% & 386 \\
21\,cm Tidal Reconstruction ($\bar{r}_{2D}$) & 0.743 & 0.971 & 30.7\% & 342 \\
N-body baryonic displacements (test loss) & 0.613 & 0.230 & 63\% & 233 \\
\bottomrule
\end{tabular}
\caption{Summary of performance improvements across four evolutionary runs on three cosmological applications. BAO and tidal reconstruction metrics are cross-correlation coefficients (higher is better); tSZ uses L1 loss (lower is better). All evolved results represent the best algorithm discovered during the evolutionary run.}
\label{tab:overview_results}
\end{table}

\subsection{Summary of Achievements}
\label{sec:sumamryachieve}

We run MadEvolve on three problems of computational cosmology, using four evolutionary runs. We briefly summarize the results, which will be explained in detail in the following sections. 

For the application of initial condition and BAO reconstruction, our first algorithm evolution starts from the widely used standard BAO reconstruction baseline. The evolved algorithm incorporates higher-order perturbation theory corrections on top of the standard Zel'dovich displacement, and uses multiple tracer fields with scale-dependent weights. This achieves $\bar{r}_{\text{BAO}} = 0.924$ compared to 0.752 for the optimized Zel'dovich baseline, a 22.8\% improvement that translates to substantially better recovery of initial conditions at quasi-linear scales ($k \sim 0.1$--$0.3\,h\,\text{Mpc}^{-1}$) where nonlinear gravitational evolution most severely degrades the BAO signal. However, the evolution does not quite reach state-of-the-art performance. We thus ran a second evolution run starting from the stronger iterative reconstruction baseline \citep{Schmittfull2017}, which is the best human algorithm to our knowledge. In this run, MadEvolve again improves the target metric from $\bar{r}_{\text{BAO}} = 0.933$ for the iterative baseline to $\bar{r}_{\text{BAO}} = 0.959$ after evolution. This shows that combining a human-designed state-of-the-art algorithm with LLM-evolved refinements yields the best overall results. Notably, evolution starting from standard reconstruction did not independently rediscover the iterative reconstruction algorithm, despite this being a well-known technique; this suggests that the system is more effective at refining and extending existing algorithms than at retrieving known solutions from its training data.

The 21cm foreground contamination reconstruction application yielded the most striking results. Starting from a baseline cross-correlation of $\bar{r}_{2D} = 0.743$ in the wedge-contaminated region of Fourier space, the evolved algorithm achieves $\bar{r}_{2D} = 0.971$, a 30.7\% improvement. The algorithm improvements primarily involve a better treatment of the anisotropic nature of the problem, which is physically reasonable.
Interestingly, the algorithm discovered at generation 52 captures essentially all of the improvement with only 9 tunable parameters, while later evolution of up to 60 parameters does not yield significant further gains, indicating that an efficient parametrization has been found.

For the effective baryonic physics from N-body simulations application, the evolved algorithm reduces test loss by 63\% while improving cross-correlation from 0.943 to 0.969. The key change is to predict the tSZ field using a multiplicative decomposition that factorizes the Compton-$y$ prediction into an electron density term and an effective temperature term, which is physically reasonable and was not enforced in the base algorithm. 

In summary, the found algorithms contain many physically plausible ideas, and indeed increase the performance of the algorithms substantially. To achieve a similar success, human researchers would need significant domain knowledge and/or try out a large number of ideas in a time consuming process. However, one would not currently call the evolved algorithms elegant. They resemble a patchwork of ideas, not a systematic human exploration which could come with optimality proofs (in certain regimes) and other mathematical derivations. Nevertheless, the found algorithms may have practical value on real data analysis tasks given further investigation. In addition, successful algorithm modifications can provide ideas for human scientists. It also seems likely that evolved algorithms will become increasingly more sophisticated with stronger models and improved evolution systems (see Sec. \ref{sec:future_directions}).  

\subsection{Difficulties and Engineering Challenges}

We encountered several practical challenges during development.
\paragraph{Prompt engineering.} Writing good prompts proved crucial for obtaining productive LLM mutations and emerged as one of the most important factors determining evolution quality.  Effective prompts must provide detailed physical background and domain knowledge, enabling the LLM to propose modifications grounded in the underlying physics. One may wonder whether this would allow the LLM to simply recall entire existing algorithms from the literature, but we have not observed this happening in practice. We further found that merely encouraging the LLM to ``improve'' the baseline produced incremental refinements that rarely escaped local optima, whereas explicitly encouraging more radical and novel approaches led to faster discovery of qualitatively new solutions. Importantly, providing specific suggestions about which aspects to modify proved counterproductive by constraining creativity; prompts that supplied rich physical context while granting the LLM full autonomy in deciding what to change produced the most effective innovations. We also observed that without explicit constraints, LLMs tended to continuously increase parameter counts, reducing interpretability. Specifying limits such as ``maintain fewer than 10 tunable parameters'' controlled this tendency without sacrificing performance: for tidal reconstruction, the 9-parameter algorithm at generation 52 achieved $\bar{r}_{2D} = 0.97$, while subsequent 30 to 60 parameter variants improved this by less than 0.4\%. The prompts for our final runs can be found in App. \ref{sec:prompts}.

\paragraph{Evaluator design.} Careful evaluator design proved equally critical, as poorly constructed metrics readily induce reward hacking. In the BAO reconstruction application, we initially used the mean cross-correlation $\bar{r}(k)$ averaged uniformly across all scales as the fitness metric. Under this objective, evolution discovered algorithms that dramatically sacrificed large-scale reconstruction accuracy to achieve marginal gains at small scales, exploiting the equal weighting in the average. The resulting algorithms were scientifically useless despite achieving high fitness scores, as BAO analyses primarily rely on large-scale modes where the acoustic signal resides. To prevent such behavior, we introduced explicit penalty terms that constrained the minimum acceptable correlation at large scales, ensuring that improvements at small scales could not come at the expense of the primary science target. This modification effectively eliminated reward hacking and guided evolution toward algorithms with balanced performance across the relevant scale range.

\paragraph{Balancing Exploration and Exploitation.} The balance between exploration and exploitation required careful tuning throughout the evolutionary process. Setting the exploitation weight too high caused premature convergence to local optima, where the population became trapped refining minor variations of a suboptimal solution rather than discovering fundamentally better approaches. Conversely, excessive exploration wasted computational budget on syntactically invalid or obviously inferior programs, as overly aggressive mutations often destroyed the functional structure that made parent programs effective. The optimal balance appears to be problem-dependent, influenced by factors such as the ruggedness of the fitness landscape, the complexity of the algorithm being evolved, and the available computational budget.

Maintaining population diversity was essential for escaping local optima and sustaining productive evolution over hundreds of generations. The MAP-Elites grid and island model together prevented the population from collapsing to a single lineage, preserving multiple competing algorithmic strategies that could be recombined or serve as stepping stones to novel solutions. However, these mechanisms required careful calibration. Several early runs exhibited ``mode collapse'' where all islands converged to nearly identical programs, producing diminishing returns from continued evolution as the population lost the diversity necessary for meaningful exploration. Mitigating this issue required adjusting several hyperparameters described in Sections~\ref{subsec:population} and~\ref{subsec:llm_ensemble}, most importantly the migration interval $T_{\text{mig}}$, the migration fraction $\rho$, the UCB exploration coefficient $c_0$ and its decay rate $\gamma$, and the relative weights of generation modes. In practice, we found that short preliminary runs of 30--50 generations were sufficient to diagnose mode collapse or excessive exploration and to identify reasonable hyperparameter settings before committing to full evolutionary campaigns.

\subsection{Model Contributions and Global Statistics}

Table~\ref{tab:global_statistics} presents aggregate statistics across the four evolutionary runs. The experiments collectively ran 2{,}126 generations. The fraction of candidate programs that successfully compiled and produced meaningful results varied considerably across applications, ranging from 54.5\% for the tSZ prediction task to 85.7\% for tidal reconstruction, reflecting the differing complexity of maintaining syntactic validity and numerical stability across problem domains.

Beyond raw success rates, we report the \emph{improvement rate}: the fraction of all mutations (including failed ones) that produce a valid program with fitness exceeding its parent's. This metric directly quantifies how efficiently evolution converts computational effort into progress. Improvement rates range from 13.4\% for tSZ to 30.4\% for BAO reconstruction, indicating that roughly one in three to one in eight mutations advances the fitness frontier. The low improvement rate for tSZ reflects both its lower success rate and the difficulty of improving an already well-optimized differentiable pipeline built on the vmad framework.

\begin{table}[h]
\centering
\begin{tabular}{lcccc}
\toprule
\textbf{Metric} & \textbf{BAO} & \textbf{Iter.\ Recon.} & \textbf{Tidal} & \textbf{tSZ} \\
\midrule
Total generations & 1{,}165 & 386 & 342 & 233 \\
Successful programs & 897 & 323 & 293 & 127 \\
Success rate (\%) & 77.0 & 83.7 & 85.7 & 54.5 \\
Improvement rate (\%) & 30.4 & 29.8 & 24.8 & 13.4 \\
\bottomrule
\end{tabular}
\caption{Global statistics across the four evolutionary runs. Success rate indicates the fraction of LLM-generated programs that compile and execute without errors. Improvement rate indicates the fraction of all mutations that produce a valid program with higher fitness than its parent.}
\label{tab:global_statistics}
\end{table}

Each evolutionary run employed a heterogeneous ensemble of frontier language models operating in parallel across five islands, with model selection governed by an Upper Confidence Bound (UCB) bandit algorithm that dynamically balanced exploration against exploitation. Seven distinct models contributed mutations across the four runs, with different subsets used in each application depending on model availability at the time of the experiment. Table~\ref{tab:model_contributions} summarizes per-model contributions aggregated across all applications, reporting not only the number of mutations and success rate but also the improvement rate and the average number of mutations required to produce one fitness-improving program (mutations per improvement).

Gemini~3~Pro~Preview generated the most mutations overall (619 of 2,126, 29.1\%), while Gemini~3~Flash~Preview achieved both the highest success rate (89.5\%) and the highest improvement rate (32.9\%), requiring only 3.0 mutations per improvement on average. By contrast, GPT-5 and o4-mini required 4.9 and 4.6 mutations per improvement, respectively. To assess whether frontier models are substantially more effective than lightweight ones, we group models into frontier-class (Gemini~3~Pro, Gemini~2.5~Pro, GPT-5, GPT-5.2) and lightweight (Gemini~3~Flash, Gemini~2.5~Flash, o4-mini). The aggregate improvement rates are comparable: 27.9\% for frontier models versus 26.3\% for lightweight models, corresponding to 3.6 and 3.8 mutations per improvement, respectively. This suggests that lightweight models can serve as competitive evolutionary contributors despite their lower per-mutation success rates. Within the lightweight category, however, performance varies substantially: Gemini~3~Flash~Preview outperforms all other models on both metrics, while Gemini~2.5~Flash shows the lowest overall success rate (62.6\%), particularly on the tSZ task (21.9\%) where the stringent differentiability requirements of the vmad framework demand careful attention to operator compatibility.

\begin{table}[h]
\centering
\small
\begin{tabular}{lcccc}
\toprule
\textbf{Model} & $N$ & \textbf{Success (\%)} & \textbf{Impr.\ (\%)} & \textbf{Muts/Impr.} \\
\midrule
Gemini 3 Pro Preview & 619 & 77.9 & 28.9 & 3.5 \\
GPT-5.2 & 303 & 76.2 & 29.7 & 3.4 \\
o4-mini & 320 & 65.9 & 21.9 & 4.6 \\
Gemini 2.5 Pro & 291 & 82.1 & 28.9 & 3.5 \\
Gemini 3 Flash Preview & 228 & 89.5 & 32.9 & 3.0 \\
GPT-5 & 186 & 84.4 & 20.4 & 4.9 \\
Gemini 2.5 Flash & 179 & 62.6 & 25.7 & 3.9 \\
\midrule
Frontier (aggregate) & 1{,}399 & 79.3 & 27.9 & 3.6 \\
Lightweight (aggregate) & 727 & 72.5 & 26.3 & 3.8 \\
\bottomrule
\end{tabular}
\caption{Per-model mutation statistics aggregated across all four applications. $N$: total mutations generated; Success: fraction producing valid, executable programs; Impr.: fraction of all mutations yielding a fitness score exceeding the parent's; Muts/Impr.: average number of mutations needed to produce one fitness-improving program. Frontier models: Gemini~3~Pro, Gemini~2.5~Pro, GPT-5, GPT-5.2. Lightweight models: Gemini~3~Flash, Gemini~2.5~Flash, o4-mini. Not all models were used in every run.}
\label{tab:model_contributions}
\end{table}

The best-performing algorithms in each application were discovered by different models: Gemini~3~Pro~Preview produced the top-ranked BAO reconstruction ($\bar{r}_{\text{BAO}} = 0.924$, generation~1144), GPT-5.2 generated the best iterative reconstruction variant ($\bar{r}_{\text{BAO}} = 0.959$, generation~377), GPT-5 produced the best tidal reconstruction algorithm ($\bar{r}_{2D} = 0.973$, generation~321), and Gemini~2.5~Pro discovered the optimal tSZ predictor (test loss reduction of 63\%, generation~222). This diversity suggests that no single model dominates algorithmic discovery; different models contribute complementary strengths depending on the problem structure. The UCB selection mechanism adapted model utilization dynamically throughout evolution, adjusting sampling probabilities based on each model's track record of producing fitness improvements on each island.

\section{Application I: Initial Condition and BAO Reconstruction}\label{sec:app1}

In the following three sections we apply MadEvolve to three different applications in cosmology. In each case, we describe the computational task, human baseline algorithm, the novel evolved algorithm, and then put the evolved algorithm in context with existing approaches in the literature.

\subsection{Problem Formulation and Scientific Motivation }

Our first application is the reconstruction of the unobservable initial matter distribution of the universe from the observable late-time distribution. The primary scientific application for this process is to measure the baryon acoustic oscillations (BAO) more precisely. Baryon acoustic oscillations are a key cosmological probe, marking the scale where primordial sound waves froze at recombination. At low redshifts ($z \lesssim 2$), gravity drives bulk flows that blur the BAO signal and reduce its precision as a standard ruler. Reconstruction algorithms partially undo this blurring by estimating and removing the displacements caused by gravitational evolution.

We frame reconstruction as an optimization problem: find algorithms that maximize the cross-correlation between reconstructed and true initial density fields in Fourier space:

\begin{equation}
r(k) = \frac{\langle \delta_{\text{rec}}(\bm{k}) \delta_{\text{IC}}^*(\bm{k}) \rangle}{\sqrt{\langle |\delta_{\text{rec}}(\bm{k})|^2 \rangle \langle |\delta_{\text{IC}}(\bm{k})|^2 \rangle}},
\label{eq:Corr}
\end{equation}

where $\delta_{\text{IC}}$ is the initial field at $z \approx 127$ (linear regime), $\delta_{\text{rec}}$ is our reconstruction from $z = 0$ data, and brackets denote k-shell averaging. Our main goal is maximizing:

\begin{equation}
\bar{r}_{\text{BAO}} = \frac{1}{N_k} \sum_{k \in [k_{\min}, k_{\max}]} r(k)
\end{equation}

averaged over $k \in [0.01, 0.5]$ $h$ Mpc$^{-1}$ where BAO information lives. This measures how well we recover the initial field at BAO scales, covering both the acoustic peak ($k \sim 0.1$ $h$ Mpc$^{-1}$) and the damping tail toward smaller scales. 

\subsection{Experimental Setup}

Our evolutionary framework begins with two different reconstruction methods: the standard reconstruction \cite{eisenstein2007improving} and iterative reconstruction \citep[][]{Schmittfull2017} as the baseline algorithms. 
In the following section, we present the evolved algorithms obtained from these two baselines, respectively.

Standard reconstruction, which is widely adopted in cosmological analyses for its computational efficiency and solid theoretical foundation, employs a single-step linear displacement field estimation based on the Zel'dovich approximation. 
The reconstruction proceeds by solving for the Zel'dovich displacement field $\bm{\Psi}$ that maps particles from their observed Eulerian positions $\mathbf{x}$ back to their initial Lagrangian coordinates $\mathbf{q}$:
\begin{equation}
\mathbf{q} = \mathbf{x} - \bm{\Psi}(\mathbf{x}), \quad \bm{\Psi} = -\nabla \Phi, \quad \nabla^2 \Phi = \delta_s
\end{equation}
Here $\delta_s$ represents the smoothed overdensity field obtained by convolving the observed density contrast with a Gaussian kernel $W(k) = \exp(-k^2 R_s^2/2)$, $\Phi$ refers to the potential field.
We start with a smoothing scale of $R_s = 10$ $h^{-1}$ Mpc, which effectively suppresses small-scale nonlinear effects while preserving the large-scale velocity flow patterns crucial for BAO reconstruction. 

Built upon the standard reconstruction, the iterative baseline aims to more accurately approximate the nonlinear displacement field by repeatedly applying the linear reconstruction procedure. 
At each step, the Zel'dovich displacement field is computed from the density field using a progressively decreasing smoothing scale given by 
$R_i = 0.5^{i} \times 10\,h^{-1}\,\mathrm{Mpc},
$
with a minimum smoothing scale of $R_{\min} = 1\,h^{-1}\,\mathrm{Mpc}$. 
The cumulative displacement from the original catalog to the nearly uniform configuration provides an estimate of the Eulerian-to-Lagrangian displacement field $\bm{\chi}$. 
The reconstructed density field is then approximated as $\nabla \cdot \bm{\chi}$ according to the Eulerian continuity equation. To our knowledge, this algorithm is the human state-of-the-art algorithm (SotA), when considering only reconstruction algorithms rather than forward modeling or neural networks (which are much harder to deploy in practice). Unfortunately, unlike in machine learning, in physics there are usually no algorithm leader boards on unified data sets, which would clearly identify the SotA algorithm.

To enable the autodiff-based parameter optimization strategy described in Section~\ref{subsec:inner_opt}, we implement the entire reconstruction pipeline in JAX, ensuring that all operations---Gaussian filtering, displacement field computation, particle shifting, and Cloud-in-Cell mass assignment---are fully differentiable. Each evolved program declares its tunable parameters through structured annotations specifying initial values, valid bounds, and the optimization method. Before fitness evaluation, the system invokes JAX's reverse-mode automatic differentiation to compute gradients of $\bar{r}_{\text{BAO}}$ with respect to all continuous parameters, then applies the Adam optimizer with learning rate $\eta = 0.1$ and standard momentum coefficients ($\beta_1 = 0.9$, $\beta_2 = 0.999$). Optimization typically converges within 30--50 iterations, with early stopping triggered when the gradient norm falls below $10^{-5}$. This approach enables efficient navigation of parameter spaces containing 10 or more continuous variables, which would be prohibitively expensive to explore via grid search or derivative-free methods. The optimized parameter values are then injected back into the program source code, ensuring that subsequent fitness evaluations and offspring generation reflect the algorithm's best achievable performance.

We evaluate all candidate algorithms based on the Quijote N-body simulation suite \cite{Villaescusa2020}, which provides both final matter distributions and the corresponding initial conditions necessary for ground-truth comparison. Table~\ref{tab:quijote_params} summarizes the simulation parameters. 
The availability of true initial conditions at $z_{\text{IC}} = 127$ enables direct quantification of reconstruction fidelity through Fourier-space cross-correlation. We use Equation~\ref{eq:Corr} as the primary performance metric. Specifically, we use the BAO-averaged correlation $\bar{r}_{\text{BAO}}$, computed as the mean of $r(k)$ over 50 logarithmically-spaced bins spanning $k \in [0.01, 0.5]$ $h$ Mpc$^{-1}$.
This range encompasses both the typical BAO range at around $k \sim [0.01, 0.3]$ $h$ Mpc$^{-1}$ and the damping tail toward smaller scales where nonlinear information resides.

\begin{table}[h]
\centering
\begin{tabular}{ll}
\toprule
\textbf{Parameter} & \textbf{Value} \\
\midrule
Box size & $L  = 1000$ $h^{-1}$ Mpc \\
Particle count & $N_p = 512^3$ \\
Mass resolution & $m_p = 6.56 \times 10^{11}$ $h^{-1} M_{\odot}$ \\
Grid resolution  & $N_{\text{mesh}} = 256^3$ \\
Initial / final redshift & $z_{\text{IC}} = 127$, $z_{\text{final}} = 0$ \\
Cosmology & $\Omega_m = 0.3175$, $h = 0.6711$, $\sigma_8 = 0.834$ \\
\bottomrule
\end{tabular}
\caption{Quijote simulation parameters for BAO reconstruction evaluation.}
\label{tab:quijote_params}
\end{table}

For computational efficiency during the evolutionary search, we use the grid resolution with $256^3$ voxels, preserving the BAO features while reducing evaluation time. 
From the 100 independent realizations in the Quijote fiducial set, we use a single simulation (index 0) for fitness evaluation during evolution, with the remaining simulations reserved for final generalization testing. Even though we use only a single training simulation, we find no evidence of overfitting when applying the method to independent test simulations in Sec. \ref{sec:appli1_performance}. This design choice prioritizes rapid iteration over per-generation statistical averaging, as the autodiff-based parameter optimization already ensures each candidate algorithm achieves its optimal performance on the training simulation. 
A single fitness evaluation comprises the reconstruction algorithm execution plus up to 20 Adam iterations for parameter optimization. 
The baseline Zel'dovich method executes in approximately 10 seconds per reconstruction, while the fully-evolved algorithm with deformation tensor corrections requires approximately 76 seconds due to the additional FFT operations for computing tensor invariants and the multi-pass architecture. 
Including the autodiff optimization loop, a complete fitness evaluation ranges from 3--5 minutes for simple algorithms to 25--30 minutes for the most complex evolved variants.

\subsection{Results - Starting from Standard Reconstruction}

In the present section we present the algorithm evolution starting from a well-known but suboptimal human baseline, standard BAO reconstruction. In the following section, we present the algorithm evolution starting from the human SotA algorithm, iterative initial condition reconstruction.

\subsubsection{Evolution Dynamics and Performance Trajectory}

\begin{figure}[htbp]
    \centering
    \includegraphics[width=0.85\linewidth]{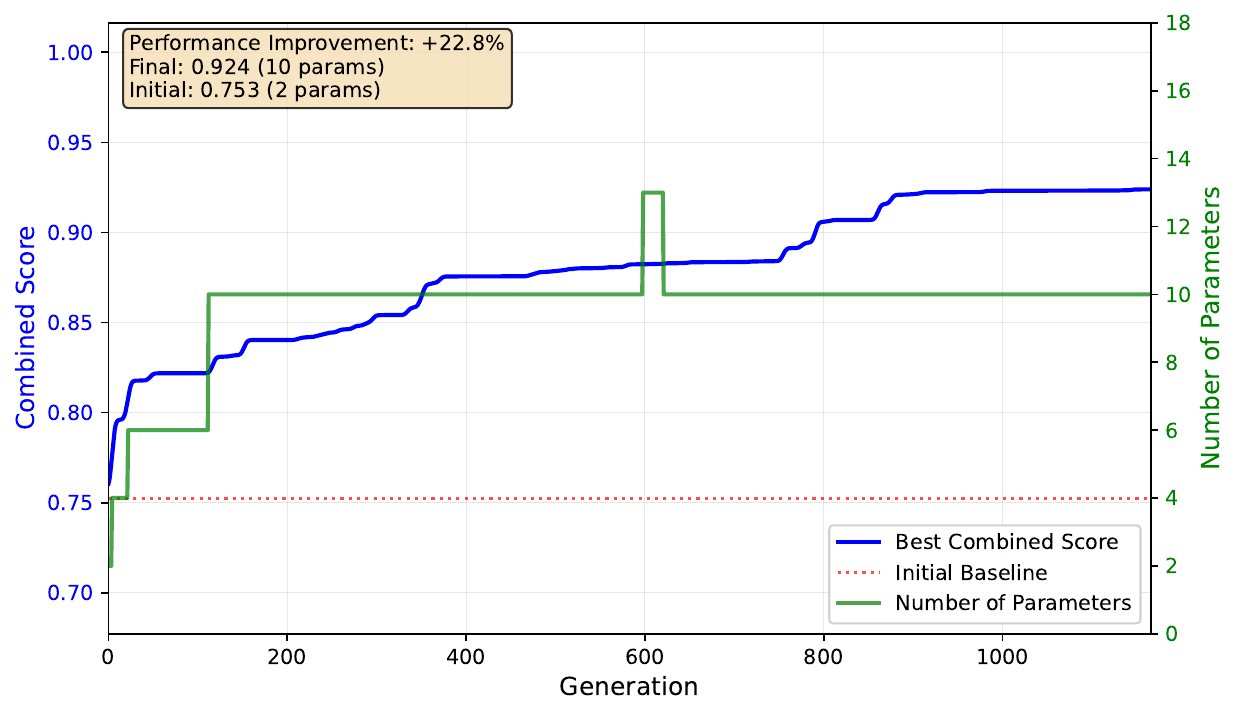}
    \caption{Evolution dynamics of BAO reconstruction algorithm discovery over 1,165 generations. The blue curve tracks the best training-set score $\bar{r}_{\text{BAO}}$ achieved by any program in the population, while the green curve indicates the number of autodiff-optimizable parameters in the best-performing algorithm. The red dashed line marks the initial baseline performance ($\bar{r}_{\text{BAO}} = 0.752$) achieved by the two-parameter Zel'dovich method. The evolutionary trajectory exhibits distinct phases: rapid initial improvement during the first 100 generations, a gradual increase through $\bar{r}_{\text{BAO}} \approx 0.85$--$0.88$ as parameter count stabilizes near 10, a notable jump around generation 600--650, and continued refinement toward the final best score of $\bar{r}_{\text{BAO}} = 0.924$ found at generation 1,144. The +22.8\% improvement demonstrates sustained evolutionary progress through the combination of LLM-proposed structural modifications and autodiff parameter optimization.}
    \label{fig:bao_evolution}
\end{figure}

We applied MadEvolve to standard BAO reconstruction as described above. The evolutionary search discovered progressively more sophisticated reconstruction algorithms through a combination of structural innovations proposed by the LLM (outer optimization loop) and automatic parameter optimization via autodiff (inner optimization loop). Figure~\ref{fig:bao_evolution} illustrates the progression of the best-performing algorithm over 1,165 generations. Starting from the baseline standard reconstruction with $\bar{r}_{\text{BAO}} = 0.752$ and only two tunable parameters (smoothing scale and bias), the system explored modifications including higher-order perturbative corrections and spectral refinement techniques.

During the first 100 generations, performance improved rapidly from 0.752 to approximately 0.82. The parameter count then increased to approximately 6 as physics-based corrections were introduced, and performance continued climbing steadily through $\bar{r}_{\text{BAO}} \approx 0.85$--$0.88$. A significant structural breakthrough occurred around generations 600--650, when the system discovered three key innovations: \emph{tensor invariant augmentation}, \emph{spectral de-warping}, and \emph{coherence-gated fusion}. We discuss the physical interpretation of these innovations in the next section. The final evolved algorithm achieves $\bar{r}_{\text{BAO}} = 0.924$, representing a 22.8\% improvement over the optimized baseline, with the best solution found at generation 1,144.

\subsubsection{Discovered Algorithmic Techniques}

We start with MadEvolve's own LLM-generated report of its innovations, which has been lightly edited for conciseness.
We then provide comments by a human expert who reviewed the report and final code.
For the complete unedited report please refer to our website \footnote{\href{https://madevolve.org/reports/bao_evolution_report.pdf}{https://madevolve.org/reports/bao\_evolution\_report.pdf}}
. The excerpt below is contained in the section \emph{Best Evolved Algorithm Analysis}.

\begin{llmreport}
This algorithm employs a \emph{hybrid iterative displacement field reconstruction} method, rooted in the Zeldovich approximation but significantly augmented by higher-order Lagrangian perturbation theory (LPT) terms and a multi-stage spectral filtering process. It operates in two distinct phases: a primary bulk-flow removal step (standard reconstruction) followed by a sophisticated ``physics augmentation'' phase that attempts to recover information lost to non-linear evolution using a basis of tidal and density invariants.

The reconstruction begins with a standard Zeldovich step, solving for the displacement field $\Psi$ using a smoothed density field and removing redshift-space distortions (RSD) via an anisotropic kernel. The innovation lies in the subsequent steps:
\begin{enumerate}
\item \emph{Tensor Invariant Augmentation:} The algorithm computes the shear tensor invariants ($I_1$, $I_2$, $I_3$) of the initial displacement field and uses them, along with an advection term ($\Psi \cdot \nabla \delta$), to construct a secondary displacement source. This mimics 2LPT (Second-order Lagrangian Perturbation Theory) corrections but with free, evolved coefficients (\texttt{gamma\_adv}, \texttt{gamma\_I2}, etc.) rather than fixed theoretical values.
\item \emph{Spectral De-Warping (Stage A):} A ``Tri-Basis De-Warping'' step attempts to fit the high-$k$ residual density field using a basis set composed of $\delta^2$, $|\nabla \delta|^2$, and tidal scalar $s^2$. This acts as a bias renormalization scheme, correcting for non-linear mode coupling at smaller scales.
\item \emph{Coherence-Gated Fusion (Stage B):} The final density field is constructed by fusing the de-warped field with a set of tracers (log-density, non-linear density, scale-dependent bias terms). This fusion is governed by a spectral coherence gate, which only allows the mixing of terms that show strong cross-correlation with the reference field in specific $k$-bands.
\end{enumerate}

The most distinct departure from standard reconstruction is the \emph{differentiable, band-wise spectral fusion}. Standard methods usually apply a global smoothing kernel. This algorithm, however, uses \texttt{jax.lax.scan} to iterate through specific wavenumber bands ($k \sim 0.2$--$0.6\,h\,\text{Mpc}^{-1}$), applying a Wiener-like filter that adaptively blends the reconstructed field with non-linear tracers based on their local spectral coherence. Additionally, the explicit inclusion of a Gram-Schmidt orthogonalization step between the primary and secondary displacement channels ensures that the higher-order corrections do not re-introduce large-scale bulk flows already handled by the Zeldovich step.

\end{llmreport}

As a general observation, in this report and the others presented below, the LLMs are perhaps overly creative in their choice of names for their inventions. Terms like \enquote{spectral de-warping} or \enquote{Coherence-Gated Fusion} are not standard in cosmology, and do not even seem to appear on the internet. Nevertheless, as we will discuss, the presented techniques are often reasonable and indeed must improve the target metric to enter the report. 

The first innovation focuses on incorporating a range of higher-order terms. 
By integrating these contributions directly into the displacement calculation, the algorithm is able to accurately capture nonlinear mode coupling, which naturally leads to improved reconstruction performance, as is shown by previous work \citep[e.g.][]{Tassev:2012hu, Achitouv:2015gma, Schmittfull2017}.

The second innovation addresses a common concern in reconstruction: the reconstructed field remains partially non-Gaussian due to residual nonlinearities. 
By explicitly modeling and subtracting these small-scale non-Gaussian contributions, the algorithm effectively reduces the nonlinearity of the field, thereby bringing it closer to the Gaussian initial conditions.

The third innovation has not been explored in previous reconstruction algorithms. 
In this step, the algorithm combines multiple tracer fields derived from the reconstructed density, such as $\delta_{\mathrm{rec}}$, $\log(1 + \delta_{\mathrm{rec}})$, and $\nabla \cdot \bm{\Psi}$, using scale-dependent weights $w_i$ to construct the final result. 
These weights are determined so that components that are more strongly correlated with the target signal are assigned higher weights. 
With an appropriate weighting scheme, this approach can effectively enhance the signal while suppressing noise, leading to improved reconstruction performance.

We note that in the newly discovered algorithm, the LLM introduces the anisotropic smoothing.
According to the generated report, this is designed to address redshift-space distortion (RSD) features, even though such effects are not expected given that the simulations are performed in real space. 
This design choice appears to be motivated by insights drawn from existing literature. 
In practice, however, the learned smoothing scales along different directions are found to be very similar, resulting in an effectively isotropic smoothing kernel. 
Consequently, the method remains fully consistent with the real-space nature of the simulations and does not introduce spurious anisotropies in the reconstruction results.
Moreover, this feature may prove beneficial when extending the algorithm to applications in the redshift space. However, this is a case where additional prompting (explaining that the simulations are isotropic) would have been beneficial. We also note that our single training simulation may have been slightly anisotropic (due to its random initial conditions), however we found no significant performance gain or degradation when we manually removed this feature on the test simulations.

\subsubsection{Performance Comparison}
\label{sec:appli1_performance}

Table \ref{tab:performance_detailed} compares the baseline and evolved reconstruction methods, evaluated on 9 held-out test simulations not used during training.
The evaluation uses the cross-correlation coefficient $r(k)$ averaged over the BAO-relevant range $k \in [0.01, 0.5]\,h\,\text{Mpc}^{-1}$. 
The evolved algorithm achieves an improvement of $22.8\%$ at the cost of approximately eight times more computation time.

\begin{figure}[htbp]
    \centering
    \includegraphics[width=0.99\linewidth]{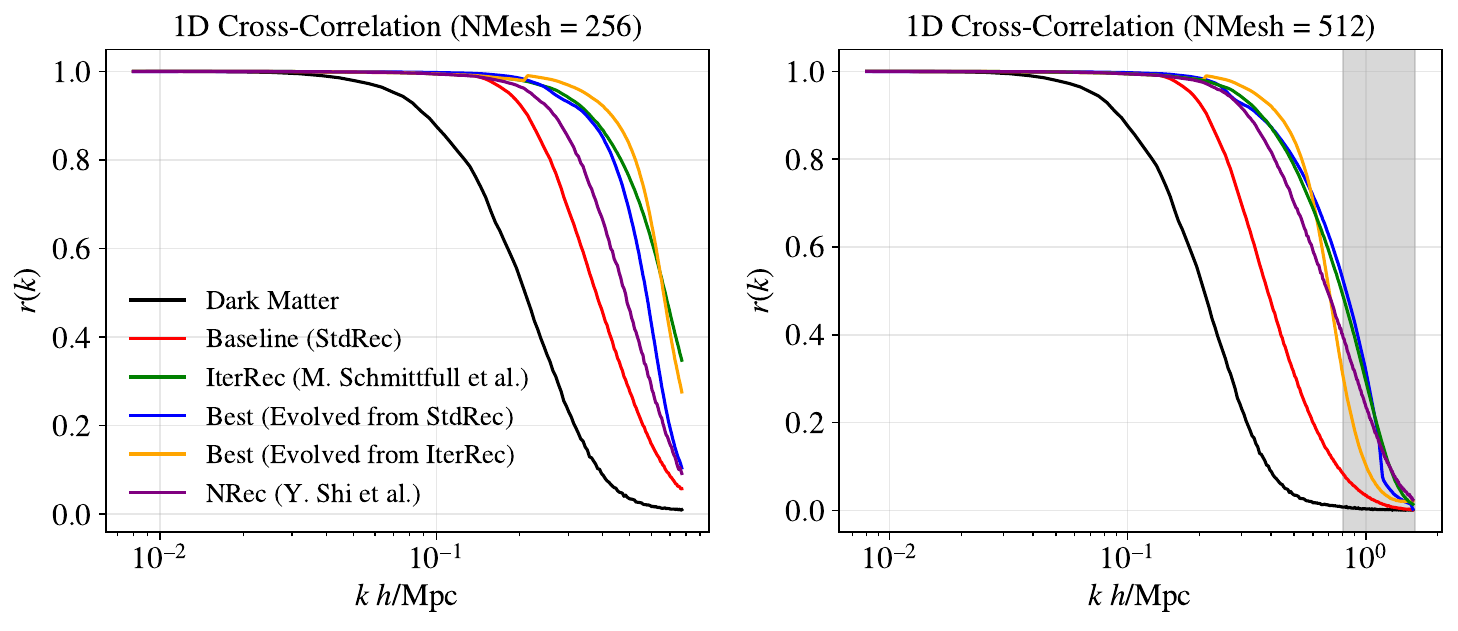}
    \caption{A comparison of reconstruction performance for different algorithms.
    The results are averaged over nine simulations in the test set ({\tt fiducial 1-9}).
    The curves shows the cross-correlation function of dark matter field (black), standard reconstruction result (red), iterative reconstruction result (green), evolved algorithm result based on standard reconstruction (blue), evolved algorithm result based on iterative reconstruction (orange), and non-linear reconstruction result (purple) with respect to the initial condition. \textbf{Left}: Reconstruction is executed on a $256^3$ mesh (which we evolved on). \textbf{Right}: Reconstruction is executed on a $512^3$ mesh (which we did not evolve on). The shaded region corresponds to Fourier modes beyond the Nyquist frequency in the evolving configuration.}
    \label{fig:rk_resolution}
\end{figure}

Upon the averaged $r(k)$, in figure \ref{fig:rk_resolution} we show the cross-correlation coefficient $r(k)$ for both baseline (red) and evolved algorithms (blue) on different scales. 
At large scales ($k < 0.1\,h\,\text{Mpc}^{-1}$), both methods achieve $r(k) \approx 1$, indicating excellent recovery of linear modes where the Zel'dovich approximation is accurate. 
The evolved algorithm maintains slightly higher correlation even in this regime, with the mean large-scale $r(k)$ improving from 0.988 to 0.996 over $k \in [0.01, 0.2]\,h\,\text{Mpc}^{-1}$. 
The algorithms diverge significantly at intermediate scales ($0.1 < k < 0.3\,h\,\text{Mpc}^{-1}$), where the evolved hybrid augmented method maintains $r(k) > 0.95$ while the baseline drops below $r(k) \approx 0.8$. 
Notably, at $k = 0.188\,h\,\text{Mpc}^{-1}$, the evolved algorithm achieves $r(k) = 0.987$ compared to $r(k) = 0.953$ for the baseline---a 3.5\% improvement at a critical quasi-linear scale where BAO information is concentrated. 
At smaller scales ($k > 0.3\,h\,\text{Mpc}^{-1}$), both methods show declining correlation as fully nonlinear structure formation dominates, but the evolved algorithm degrades far more gradually. 
The maximum improvement reaches approximately 0.40 near $k \approx 0.4\,h\,\text{Mpc}^{-1}$, demonstrating that the evolved algorithm successfully recovers information from the quasi-linear and mildly nonlinear regimes that standard reconstruction cannot access. 

In the right panel of Figure~\ref{fig:rk_resolution}, we further evaluate the performance of different algorithms at a higher resolution. 
The discovered algorithm remains consistently better than the standard reconstruction on the finer mesh, even extending to scales smaller than the original Nyquist frequency of the $256^3$ grid. 
This behavior suggests that MadEvolve captures physically meaningful improvements, rather than merely benefiting from numerical fitting or resolution-dependent effects, providing an advantage over traditional machine-learning approaches.

\begin{table}[h]
\centering
\begin{tabular}{lcccc}
\toprule
\textbf{Algorithm} & \textbf{$\bar{r}_{\text{BAO}}$} & \textbf{Correlation} & \textbf{Runtime} & \textbf{Parameters} \\
& \textbf{($k \in [0.01, 0.5]$)} & \textbf{(real space)} & \textbf{(seconds)} & \\
\midrule
Baseline Zel'dovich (autodiff) & 0.752 & 0.344 & 9.7 & 2 \\
Evolved: Hybrid Augmented & \textbf{0.924} & \textbf{0.581} & 76.1 & 10 \\
\midrule
\textbf{Improvement} & \textbf{+22.8\%} & \textbf{+68.9\%} & -- & -- \\
\bottomrule
\end{tabular}
\caption{Performance comparison of BAO reconstruction algorithms evaluated on 9 held-out test simulations.
}
\label{tab:performance_detailed}
\end{table}

\subsection{Results - Starting from Iterative Reconstruction}

As we discuss further below, our evolution starting from the standard reconstruction algorithm did not quite reach the performance of the human SotA algorithm, iterative reconstruction. In our second evolution run, we evolved starting from this SotA algorithm.

\subsubsection{Evolution Dynamics and Performance Trajectory}

\begin{figure}[htbp]
    \centering
    \includegraphics[width=0.85\linewidth]{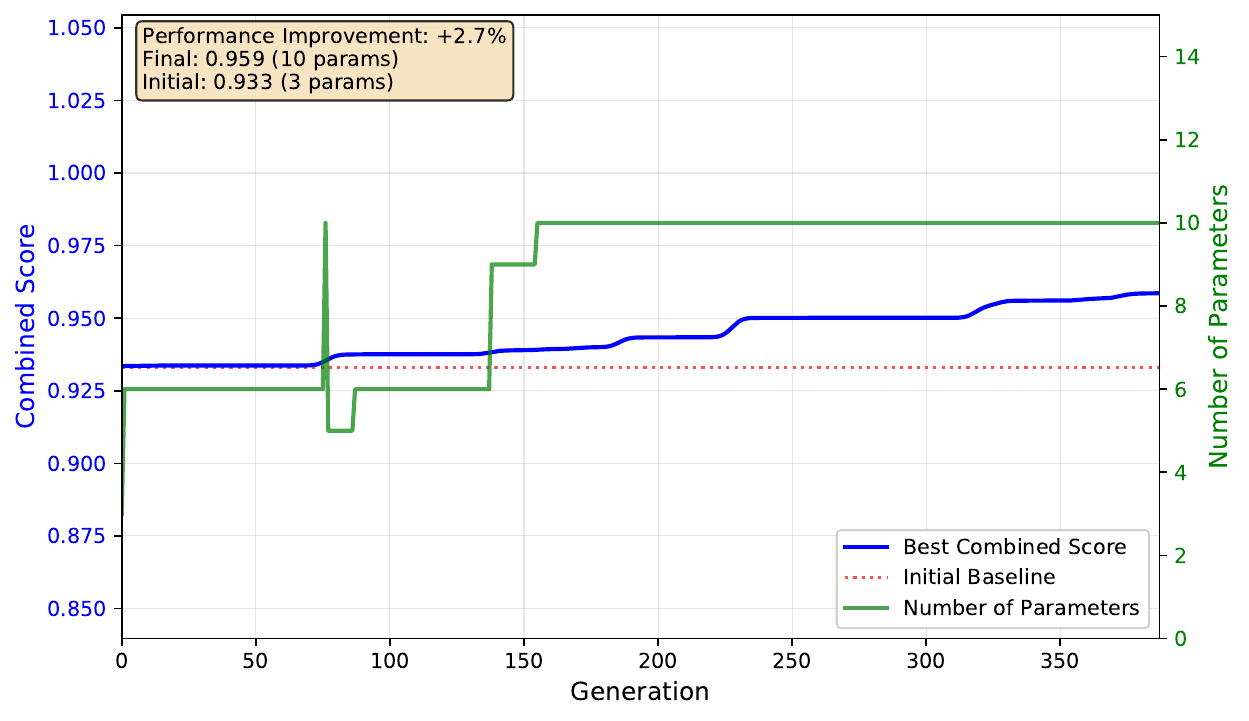}
    \caption{Evolution dynamics for the BAO reconstruction run starting from the iterative reconstruction baseline \citep{Schmittfull2017}. The blue curve tracks the best $\bar{r}_{\text{BAO}}$ achieved by any program in the population across 386 generations. Starting from $\bar{r}_{\text{BAO}} = 0.933$, the evolutionary search discovers differentiable refinements that push performance to $\bar{r}_{\text{BAO}} = 0.959$ ($+2.8\%$), demonstrating that the framework can improve upon an already well-optimized reconstruction algorithm.}
    \label{fig:iterative_bao_evolution}
\end{figure}

To further probe the potential of MadEvolve to improve over well-optimized SotA algorithms, we conducted a separate evolution run using the iterative reconstruction method \citep{Schmittfull2017} as the starting algorithm, rather than the single-step Zel'dovich method. 
The iterative baseline, which applies eight successive displacement corrections with geometrically decreasing smoothing scales ($R = 0.5^{i}\times 10\,h^{-1}\,\text{Mpc}$, $R_{\min}=1\,h^{-1}\,\text{Mpc}$), already achieves $\bar{r}_{\text{BAO}} = 0.933$ on the same evaluation setup---substantially above the Zel'dovich baseline of $0.752$.
Starting from this stronger baseline, the evolutionary search still discovers meaningful improvements: after 386 generations, the best evolved algorithm reaches $\bar{r}_{\text{BAO}} = 0.959$, a $+2.8\%$ improvement over the iterative baseline (Fig.~\ref{fig:iterative_bao_evolution}).
The discovered refinements---including \emph{guided anisotropic diffusion}, \emph{residual advection}, and \emph{2LPT source term}---are applied as differentiable post-processing stages on top of the non-differentiable iterative core. We discuss the physical interpretation of these innovations in the next section. While the relative improvement is expectedly smaller than the $+22.8\%$ gain achieved from the Zel'dovich starting point, this result confirms that the framework can extract additional physical information even when the baseline is already close to optimal.

\subsubsection{Discovered Algorithmic Techniques}
We start with MadEvolve's own LLM-generated report of its innovations, which has been lightly edited for conciseness.
We then provide comments by a human expert who reviewed the report and final code.
For the complete unedited report please refer to our website \footnote{\href{https://madevolve.org/reports/iterative_evolution_report.pdf}{https://madevolve.org/reports/iterative\_evolution\_report.pdf}}
. The excerpt below is contained in the section \emph{Best Evolved Algorithm Analysis}.

\begin{llmreport}
The algorithm employs a hybrid ``Shear-Aware + EFT Crossover'' strategy. It begins with a standard, non-differentiable iterative particle reconstruction (an iterative Zeldovich approximation) to establish a robust baseline density field. This baseline is then refined using a differentiable, grid-based post-processing step. This second stage treats the reconstruction error as a fluid dynamical problem, applying corrections derived from Effective Field Theory (EFT) and large-scale environmental tensors to recover information lost to non-linear evolution in the high-$k$ regime.

The refinement stage is physically motivated by the need to correct for mode-coupling and stream-crossing effects that standard displacement methods miss. The algorithm constructs a ``modulation weight'' ($w_{\mathrm{mod}}$) based on the large-scale environment (density and tidal shear), effectively identifying regions where non-linearities are strongest. It then applies three primary physical corrections modulated by this weight:
\begin{enumerate}
\item \emph{Guided Anisotropic Diffusion:} A diffusive term ($\gamma_{\mathrm{diff}}$) directed by the tidal tensor, smoothing the field along filamentary structures rather than isotropically.
\item \emph{Residual Advection:} It advects the ``residual'' density (small-scale fluctuations) along the large-scale velocity flows, correcting for small-scale displacements relative to the bulk flow.
\item \emph{2LPT Source Term:} A second-order Lagrangian Perturbation Theory source term ($\beta_{\mathrm{2lpt}}$) is injected to capture non-Gaussian features generated by gravity. Additionally, it includes an isotropic EFT counterterm ($c_{s}^2 k^2 \delta$) to regularize small-scale power and a ``phase nudging'' step that gently rotates Fourier phases toward those of an arcsinh-compressed reference field, acting as a soft non-Gaussian prior.
\end{enumerate}

The most innovative aspect is the \emph{strict spectral gating combined with environment-modulated residuals}. Unlike standard methods that apply corrections globally, this algorithm calculates a ``gate'' that strictly forbids any modification to modes $k < 0.21\,h\,\text{Mpc}^{-1}$. This ensures the linear regime---which is already well-recovered by the baseline method---is perfectly preserved. Furthermore, the coupling of the shear invariant ($s^2$) to the modulation weight allows the algorithm to aggressively target high-density, high-shear regions (knots and filaments) for correction while leaving voids relatively untouched, a nuance rarely seen in standard EFT-based reconstruction.

\end{llmreport}

MadEvolve only applies modest corrections on small scales to the iterative reconstruction method.
On large scales ($k < 0.21 \ \mathrm{h/Mpc}$), it directly preserves the result of iterative reconstruction, which is reasonable as iterative reconstruction is already known to perform near optimally in this regime for dark matter fields.
On small scales, the evolved algorithm incorporates a range of higher-order perturbation theory inspired terms and other nonlinear terms (guided anisotropic diffusion term and residual advection term) to refine the reconstruction, with their coefficients optimized through automatic differentiation.
With the increased flexibility of these fitted corrections, the method achieves improved performance. The LLM explanations are drawing from a large number of concepts used in cosmology, but the creativity (or randomness) in their combination makes the resulting algorithm difficult to interpret in detail by a human expert.

\subsubsection{Performance Comparison}
Table~\ref{tab:performance_iterative} summarizes the quantitative comparison between the iterative reconstruction baseline and the evolved algorithm, evaluated on 9 held-out test simulations not used during training.
The evolved algorithm achieves a $2.8\%$ improvement in the mean BAO metric $\bar{r}_{\text{BAO}}$ at a modest $13.4\%$ increase in computation time.

In Figure~\ref{fig:rk_resolution}, we present the cross-correlation coefficient $r(k)$ for both the iterative reconstruction (green) and the evolved algorithm (orange) across different scales.
On large scales ($k < 0.21\,h\,\mathrm{Mpc}^{-1}$), the evolved algorithm preserves the result of the iterative reconstruction, leading to identical performance in this regime.
On smaller scales ($k > 0.21\,h\,\mathrm{Mpc}^{-1}$), the evolved algorithm generally outperforms the baseline, indicating that the additional nonlinear corrections effectively recover information lost to mode coupling.
However, the performance gain diminishes as one approaches the Nyquist frequency ($k \gtrsim 0.6\,h\,\mathrm{Mpc}^{-1}$), where both methods are limited by resolution and residual small-scale nonlinearities.

In the right panel of Figure~\ref{fig:rk_resolution}, we further evaluate the performance of different algorithms at a higher resolution. 
Without re-optimizing the free parameters, the evolved algorithm shows improved performance on quasi-linear scales, but falls behind the iterative reconstruction on smaller scales, including those approaching and exceeding the Nyquist frequency. 
This behavior is expected, as the evolutionary process is constrained by the resolution of the $256^3$ grid and cannot accurately capture small-scale dynamics beyond this limit during evolution. 
Such limitations could potentially be mitigated by performing the evolutionary search on a finer mesh. 
We leave this for future work.

\begin{table}[h]
\centering
\begin{tabular}{lccc}
\toprule
\textbf{Algorithm} & \textbf{$\bar{r}_{\text{BAO}}$} & \textbf{Correlation} & \textbf{Runtime} \\
& \textbf{($k \in [0.01, 0.5]$)} & \textbf{(real space)} & \textbf{(seconds)} \\
\midrule
Iterative Reconstruction & $0.933 $ & $\mathbf{0.660}$ & 32.7 \\
Evolved: Iterative Augmented & $\mathbf{0.959 }$ & $0.639 $ & 37.1 \\
\midrule
\textbf{Improvement} & \textbf{+2.8\%} & $-3.3\%$ & +13.4\% \\
\bottomrule
\end{tabular}
\caption{Performance comparison of evolved algorithm against iterative BAO reconstruction, evaluated on 9 held-out test simulations.}
\label{tab:performance_iterative}
\end{table}

\subsection{Discussion}

Our results demonstrate that the evolutionary framework is able to discover reconstruction programs that substantially increase the cross-correlation with the true initial conditions on BAO-relevant scales, yielding a significant improvement over the input reconstruction algorithm. We note that these results were obtained on simulations of the matter distribution and do not directly imply gains on real galaxy surveys, however this additional realism is not the subject of the present work.

It is important to note that standard reconstruction and iterative reconstruction are two of numerous reconstruction methods. 
Beyond these two methods, substantial effort has been devoted to improving upon the Zel’dovich approximation. These developments include (i) estimating non-linear displacement fields from the density field using higher-order schemes \citep[e.g.][]{Tassev:2012hu, Achitouv:2015gma, 2017PhRvD..96l3502Z, 2018PhRvD..97b3505S}, for which we also implement the method of \citet[][]{2018PhRvD..97b3505S} in Figure~\ref{fig:rk_resolution} for comparison (purple), (ii) machine-learning-based approaches such as convolutional neural networks \citep[e.g.][]{2021MNRAS.501.1499M, 2023MNRAS.520.6256S, 2023MNRAS.523.6272C, 2025JCAP...09..039P}, and (iii) forward-model or Bayesian frameworks that infer initial conditions by fitting the full non-linear mapping between initial and final density fields, often referred to as field-level inference \citep[e.g.][]{2013MNRAS.432..894J, 2019A&A...625A..64J, 2019A&A...621A..69R, 2019JCAP...01..042S, 2020JCAP...11..008S, 2021JCAP...03..058N, 2023JCAP...07..063K, 2023arXiv230709504B, 2024PhRvL.133v1006N}. 

Within this broader landscape, the present results indicate that the evolutionary framework has the potential to further refine existing reconstruction schemes. 
By systematically exploring modifications to established algorithms, the search process can identify combinations of physically motivated ingredients that enhance performance beyond their original formulations. 
This suggests that automated evolutionary discovery may serve as a complementary tool for improving and extending current reconstruction methodologies.

At the same time, we note that the performance also depends on the choice of the input algorithm. 
We conclude that the best results are achieved when combining human-designed state-of-the-art methods with LLM-guided refinement, rather than relying solely on LLM-generated algorithms. Interestingly, despite their large literature knowledge, LLMs appear not to be able to simply recall the entire state-of-the-art algorithm from memory (i.e. here when we evolve from standard reconstruction, the LLM does not propose the iterative algorithm). Setting a strong human baseline thus remains essential for optimal performance.

\section{Application II: 21cm Foreground Contamination Reconstruction}\label{sec:app2}

Our second application, a different reconstruction task, is somewhat less well-studied than the first, but is rapidly gaining importance with the advent of 21cm interferometry surveys.

\subsection{Problem Formulation and Scientific Motivation}

Observations of the 21cm hyperfine transition of neutral hydrogen at cosmic dawn and reionization ($z \sim 6$--$30$) offer a powerful probe of early universe physics. However, foreground emission from our Galaxy and extragalactic radio sources is several orders of magnitude brighter than the cosmological signal. These foregrounds have smooth spectra, making them separable in principle, but imperfect calibration and the frequency-dependent beam pattern of interferometric arrays cause \emph{mode mixing} that leaks foreground power into otherwise clean cosmological modes, manifested as a characteristic ``wedge'' in the cylindrical ($k_\perp, k_\parallel$) Fourier space \citep[][]{Morales:2012kf, Hazelton:2013xu, Datta:2010pk, Chapman:2014sfa}, where $k_\parallel$ is the wave-number in the direction of the line-of-sight, $k_\perp$ is the wave-number in the transverse plane.

Recent theoretical work shows that anisotropic gravitational evolution imprints the large-scale tidal field onto the \emph{shape} of small-scale structures \cite{2015arXiv151104680Z, Zhu_2018, 2022ApJ...929....5Z, Zang:2022qoj}. 
This indicates that information about modes lost in the foreground wedge can be statistically recovered from modes \emph{outside} the wedge by using this correlation. 
The reconstruction task thus becomes: given a 3D density field with a wedge-shaped region of missing or contaminated modes, estimate the true field by exploiting correlations between small-scale anisotropy and large-scale modes. 
We optimize algorithms to maximize the 2D correlation coefficient in cylindrical Fourier space:
\begin{equation}
r_{2D}(k_\perp, k_\parallel) = \frac{\langle \delta_{\text{rec}}(\mathbf{k}) \delta_{\text{true}}^*(\mathbf{k}) \rangle}{\sqrt{\langle |\delta_{\text{rec}}|^2 \rangle \langle |\delta_{\text{true}}|^2 \rangle}}\Big|_{(k_\perp, k_\parallel)},
\label{eq:rk_2Dbin}
\end{equation}
averaged over the wedge region.

\subsection{Experimental Setup}

\subsubsection{21cm neutral hydrogen intensity map generation}

A thorough treatment of foreground noise is beyond the scope of this paper. In this work, we will use the following simplified model to represent foreground noise in observation. Generally, we will consider the following three types of noises: astrophysical foreground, the receiver noise and the shortest baseline for interferometers.

\begin{enumerate}
    \item \textbf{Astrophysical foreground}: We simply use a high-pass filter along the line-of-sight to represent this noise.
    \begin{equation}
        W_{\mathrm{fs}}(k_\parallel) = 1 - e^{-k_\parallel^2 R_{\mathrm{fs}}^2/2},
    \end{equation}
    where $k_{\parallel}$ is the wavenumber in the line-of-sight, $R_{\mathrm{fs}}$ is the foreground scale. This filter removes the small $k_{\parallel}$ density modes to simulate the foreground contamination. In this paper, the foreground scale is chosen to be $R_{\mathrm{fs}} = 15 \ \mathrm{Mpc} / h$. This is an ideal case where we only lose the modes with $k_\parallel < 0.02 \ h\mathrm{Mpc}^{-1}$ \cite{Shaw_2015}.
    \item \textbf{Receiver noise}: The resolution of small-scale structure in a 21 cm survey is mainly determined by thermal noise $P_{N}$. The magnitude of this noise is around $P_N = 150 \sim 600 (\mathrm{Mpc}/h)^3$ for a HIRAX-like interferometer \cite{White_2017}. In this work, we use a white noise field with magnitude $P_N = 300 (\mathrm{Mpc}/h)^3$ to model this receiver noise.
    \item \textbf{Baseline for interferometer}: The largest angular scale that could be detected in a 21 cm survey is limited by the shortest baseline of the interferometer. In this work, we simply assume that modes with $k_\perp < 0.05 \ h/\mathrm{Mpc}$ are not detectable. This is equivalent to applying a step function $\Theta_s$ to the density field in the Fourier space,
    \begin{equation}
        \Theta_s(k) = \begin{cases}
    1, & \text{if } \ k_\perp \geq k_s  \\
    0, & \text{otherwise}.
\end{cases}
    \end{equation}
\end{enumerate}

\begin{figure}
    \centering
    \includegraphics[width=0.5\linewidth]{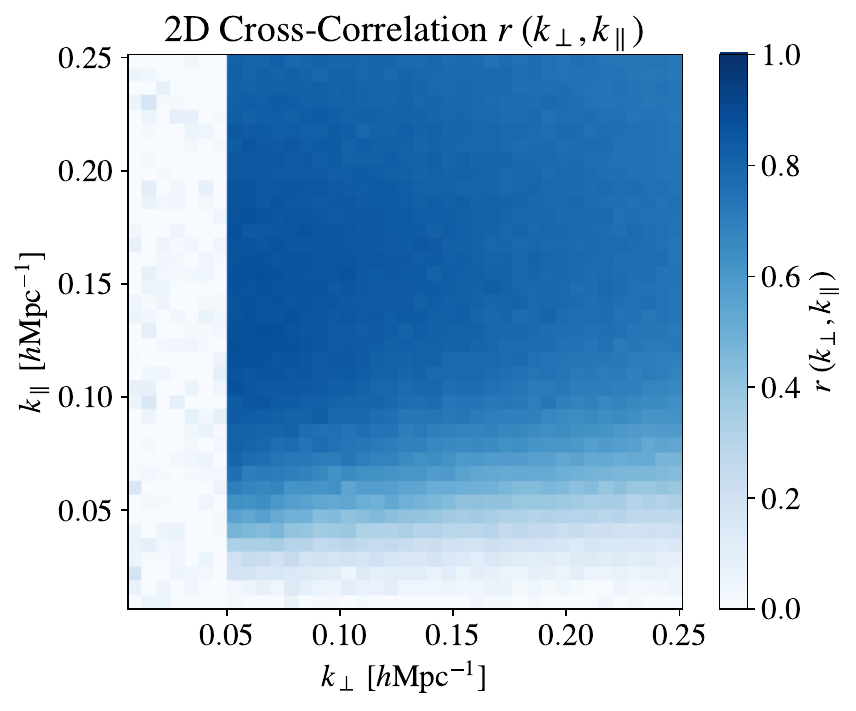}
    \caption{The cross-correlation coefficient $r(k_\perp, k_\parallel)$ of the intensity map with respect to the underlying matter field.
    The results are averaged over nine simulations in the test set ({\tt fiducial 1-9}).}
    \label{fig:r_2D_noise}
\end{figure}

In summary, given the matter density field $\delta(\bm{x})$, we generate the 21 cm neutral hydrogen field by
\begin{equation}
    \delta_{\mathrm{IM}}(\bm{x}) = \mathrm{IFFT}\left[\mathrm{FFT}\left[\delta_m(\bm{x})\right]W_{\mathrm{fs}}(k_\parallel)\Theta_s(k)\right] + N(P_N).
\end{equation}
Fig.~\ref{fig:r_2D_noise} shows the cross-correlation coefficient $r(k_\perp, k_\parallel) = P_{\delta_m\delta_{\mathrm{IM}}} / (P_{\delta_m\delta_m}P_{\delta_{\mathrm{IM}}\delta_{\mathrm{IM}}})$ of the intensity map with respect to the underlying matter field.

\subsubsection{Tidal reconstruction}
Our evolutionary framework begins with a standard implementation of the tidal reconstruction formalism \cite{Zhu_2018} as the baseline algorithm. 
This approach reconstructs the tidal shears from the observed density, calculates the tidal tensor components, and uses these components to estimate the large-scale density field. We also compare to a different quadratic estimator approach in Sec. \ref{sec:21cmqe}.

Specifically, given an observed density contrast $\delta_{\text{obs}}$ with the wedge region set to zero, the algorithm first calculates the tidal shear fields
\begin{eqnarray}
\label{eq:shearestimation}
    \hat{\epsilon}_1(\bm{x}) & = & [\delta^{w_1}(\bm{x})\delta^{w_1}(\bm{x}) - \delta^{w_2}(\bm{x})\delta^{w_2}(\bm{x})]/2, \nonumber \\
    \hat{\epsilon}_2(\bm{x}) & = & \delta^{w_1}(\bm{x})\delta^{w_2}(\bm{x}), \nonumber \\
    \hat{\epsilon}_x(\bm{x}) & = & \delta^{w_1}(\bm{x})\delta^{w_3}(\bm{x}), \nonumber \\
    \hat{\epsilon}_y(\bm{x}) & = & \delta^{w_2}(\bm{x})\delta^{w_3}(\bm{x}), \nonumber \\
    \hat{\epsilon}_z(\bm{x}) & = & [2\delta^{w_3}(\bm{x})\delta^{w_3}(\bm{x}) - \delta^{w_1}(\bm{x})\delta^{w_1}(\bm{x}) \nonumber \\
    &&- \delta^{w_2}(\bm{x})\delta^{w_2}(\bm{x})]/6,
\end{eqnarray}
where
\begin{equation}
    \label{eq:filterestimation}
    \delta^{w_j}(\bm{k})=ik_j W_R(k)\delta_{\mathrm{obs}}(\bm{k}),
\end{equation}
is the filtered gradient density field and $W_R(k) = \exp(-k^2R^2/2)$ is the Gaussian window with smoothing scale $R$.
Finally, the algorithm combines all five tidal components to estimate the full density field in Fourier space:
\begin{eqnarray}
\label{eq:5shear}
\delta_{\mathrm{rec}} =  \frac{1}{2k^2} & \left[ (k_1^2 - k_2^2)\hat{\epsilon}_1(\bm{k}) + 2k_1k_2\hat{\epsilon}_2(\bm{k}) + 2k_1k_3\hat{\epsilon}_x(\bm{k}) \right .\\ 
    &  \left . + 2k_2k_3\hat{\epsilon}_y(\bm{k}) + (2k_3^2 - k_1^2 - k_2^2)\hat{\epsilon}_z(\bm{k}) \right]. \nonumber
\end{eqnarray}
This baseline achieves moderate performance ($\bar{r}_{2D} \approx 0.74$) but struggles with nonlinear small-scale structure and fails to exploit the full information content available from anisotropic tidal correlations.

We evaluate all candidate algorithms on the Quijote N-body simulation suite \cite{Villaescusa2020}, processed to mimic 21cm observations with realistic wedge masking. 
The simulations use a box size of $L = 1000$ $h^{-1}$ Mpc with $512^3$ grid resolution. 
We apply a cylindrical mask in $(k_\perp, k_\parallel)$ space matching the expected contamination geometry from upcoming instruments such as HERA and SKA. 
All Fourier transforms use double-precision NumPy/SciPy FFT routines. 
The primary performance metric is the 2D correlation coefficient $r_{2D}(k_\perp, k_\parallel)$ computed in cylindrical bins with Equation~\ref{eq:rk_2Dbin} and averaged over the wedge region spanning $k_\perp \in [0.08, 0.65]$ $h$ Mpc$^{-1}$ and $k_\parallel < 0.08$ $h$ Mpc$^{-1}$.
We use a single simulation (sim~0) for fitness evaluation during evolution, and 9 held-out simulations (sims~1--9) for testing.

\subsection{Results}

\subsubsection{Evolution Dynamics and Performance Trajectory}

\begin{figure}[htbp]
    \centering
    \includegraphics[width=0.85\linewidth]{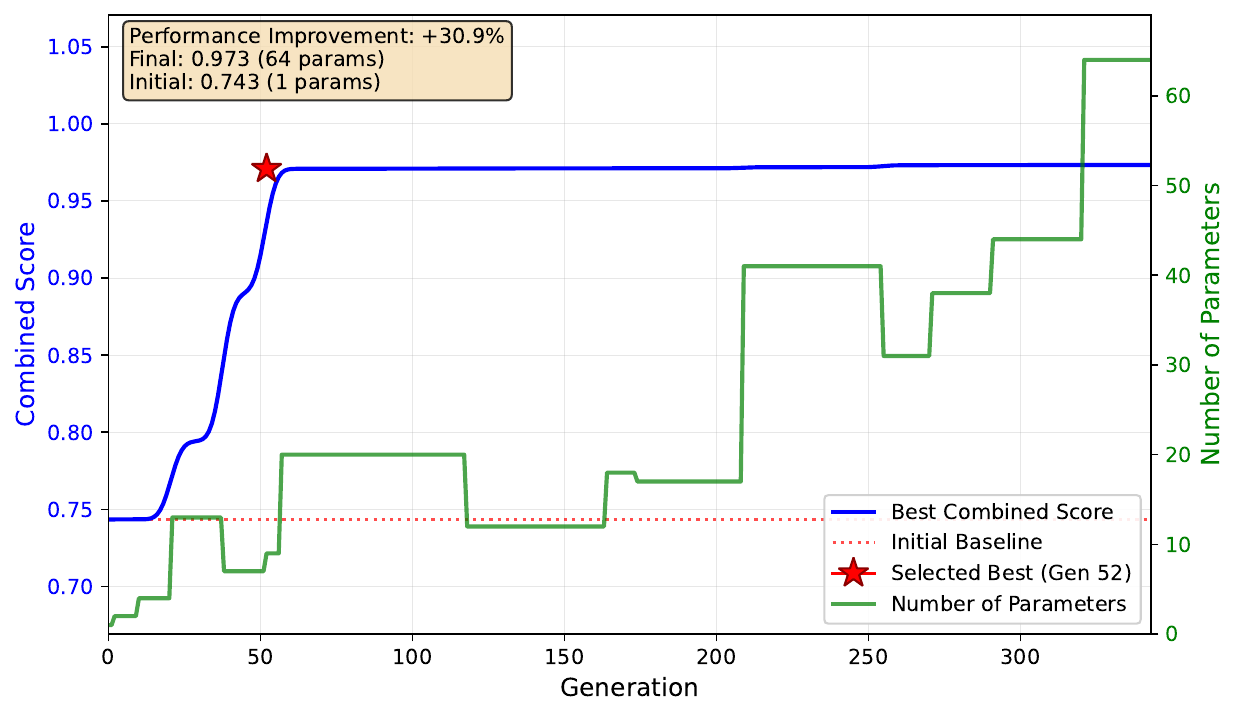}
    \caption{Evolution dynamics of tidal reconstruction algorithm discovery over 342 generations.
    The blue curve tracks the best combined score $\bar{r}_{2D}$ achieved by any program in the population, while the green curve indicates the number of tunable parameters in the best-performing algorithm. 
    The red dashed line marks the initial baseline performance ($\bar{r}_{2D} = 0.743$) achieved by the single-parameter tidal tensor method. 
    The evolutionary trajectory exhibits rapid initial improvement as the system discovers \emph{Anisotropic Filtering}, \emph{Generalized Anisotropic Potential}, and \emph{Split-Weight Tensor Projection}. 
    By generation 52, the algorithm achieves $\bar{r}_{2D} = 0.971$ with only 9 parameters, capturing the essential physics of tidal reconstruction.}
    \label{fig:tidal_evolution}
\end{figure}

Figure~\ref{fig:tidal_evolution} illustrates the progression of the best-performing algorithm over 342 generations. 
Starting from the baseline tidal tensor method with $\bar{r}_{2D} = 0.743$ and only one tunable parameter (the filter scale), the system explored modifications including \emph{Anisotropic Filtering}, \emph{Generalized Anisotropic Potential}, \emph{Split-Weight Tensor Projection}, and \emph{Component-wise Adaptive Saturation}. 
We will comment in the next section on the meaning of these innovations.
The improvement continued through generation 52, where the algorithm reached $\bar{r}_{2D} = 0.97$ with only 9 tunable parameters, demonstrating that principled architectural choices rather than extensive parameterization drive the performance gains.

The key structural breakthrough occurred during generations 30--50 when evolution discovered \emph{Split-Weight Tensor Projection} which separates the tidal tensor divergence into several components and weights them independently. 
This substantially improves reconstruction fidelity. 
Although evolution continued beyond generation 52 and discovered more complex architectures with up to 64 parameters, these additional refinements yielded only marginal improvements ($\bar{r}_{2D} = 0.973$ versus $0.97$), suggesting that the 9-parameter algorithm already captures the dominant signal recovery mechanisms. The final performance represents a 30.7\% improvement over the optimized baseline.

\subsubsection{Discovered Algorithmic Techniques}

We start with MadEvolve's own LLM-generated report of its innovations. We then provide comments by a human expert who reviewed the report and final code. 
For the complete MadEvolve report please refer to our website \footnote{\href{https://madevolve.org/reports/21cm_evolution_report.pdf}{https://madevolve.org/reports/21cm\_evolution\_report.pdf}}
. The excerpt below is contained in the section \emph{Best Evolved Algorithm Analysis}.

\begin{llmreport}
The selected algorithm implements an \emph{Anisotropic Tensor Reconstruction with Split-Component Weighting}. It improves upon the baseline through three key innovations:
\begin{enumerate}
\item \emph{Anisotropic Filtering:} Instead of a simple isotropic Gaussian, it applies a targeted high-pass filter on $k_\parallel$ modes to specifically remove foreground-contaminated modes (the ``wedge''), while preserving large-scale transverse information via Gaussian smoothing controlled by \texttt{smooth\_scale}.
\item \emph{Generalized Anisotropic Potential:} The displacement field is computed using a spectral potential with tunable spectral index and anisotropy factor ($f_{\rm ani}$), allowing the algorithm to adapt the relationship between density and displacement to the specific noise geometry of 21\,cm observations.
\item \emph{Split-Weight Tensor Projection:} The divergence of the tidal tensor is decomposed into transverse ($k_\perp^2$), longitudinal ($k_z^2$), and cross ($k_\perp k_z$) components, each weighted independently by \texttt{w\_par} and \texttt{w\_cross}. This allows the algorithm to optimally weight the geometrically distinct contributions to the reconstruction.
\end{enumerate}

Additionally, it uses \emph{Component-wise Adaptive Saturation} to prevent high-density peaks from dominating the tidal tensor, and a tunable trace weight (\texttt{w\_trace}) to control how much of the isotropic component is retained. With only 9 parameters, the algorithm achieves 99.8\% of the best-scoring 64-parameter variant's $\bar{r}_{2D}$ while running 3$\times$ faster.

\end{llmreport}

The first two innovations are specifically designed to address anisotropies in the 21 cm intensity mapping. 
These modifications are expected to improve the reconstruction performance, as foreground contamination is inherently anisotropic, while the original Gaussian filter is sub-optimal.
%This evolved algorithm introduces more detailed heuristic formalism to account for anisotropic structures, which have the potential to be directly applied in practical data analyses. 
Specifically, the anisotropic filter in the first innovation is given by
\begin{equation}
W(\mathbf{k}) = \exp\left(-\frac{k^2 R_s^2}{2}\right)
\left[1 - \exp\left(-\frac{k_\parallel^2}{2 k_{\parallel,\min}^2}\right)\right],
\end{equation}
where $k_{\parallel,\min} = 0.1863  \ h\, \mathrm{Mpc}^{-1}$ is a free parameter to be optimized, which consists of an isotropic part and an anisotropic part. 
This filter suppress the noise in the foreground wedge and keeps the small-scale uncontaminated modes.
Compared to the isotropic Gaussian filter used by the original tidal reconstruction (primarily for computational efficiency), this new anisotropic filter more closely resembles a Wiener filter, and is thus expected to provide better results.

The second innovation introduces anisotropic features in the potential calculation:
\begin{equation}
\Phi_{\mathrm{eff}}(\mathbf{k}) = - 
\frac{\delta(\mathbf{k})}
{k_\perp^2 + f_{\mathrm{ani}} \, k_\parallel^2 + k_s^2},
\end{equation}
where $f_{\mathrm{ani}} = 1.055$ and $k_s = 0.126\,h\,\mathrm{Mpc}^{-1}$ are the optimized parameters. 
According to the report, this modification is intended to mitigate RSD effects.
Although the simulations considered here are performed in real space, the parameter $f_{\mathrm{ani}}$ introduces an anisotropic response along the line of sight that can be beneficial for our task. Note that the optimized value of $f_{\mathrm{ani}}$ is close to unity, consistent with the real space setup. We also point out the unusual naming choice \enquote{spectral potential}, which seems to mean a potential expressed in Fourier space.

The third innovation combines different tidal terms with independently optimized coefficients, $w_\parallel$ and $w_{\times}$, which provides additional flexibility to adapt to different anisotropic noise configurations. 
This design again allows the algorithm to better capture direction-dependent features in the data.

The new algorithm also uses \emph{component-wise adaptive saturation}, which is designed to suppress nonlinear contributions in the field. 
When computing the quadratic tidal shear terms, highly nonlinear regions associated with dense peaks generate outliers that potentially dominate the tidal components.
This issue has been identified in previous work \citep[][]{2015arXiv151104680Z}, where it was addressed through a logarithmic transformation of the density field.
In contrast, the present algorithm introduces an adaptive damping mechanism to mitigate these extreme contributions, effectively reducing nonlinear contamination in the tidal shear fields and enabling a more accurate recovery of large-scale modes.

\subsubsection{Performance Comparison}

Table~\ref{tab:tidal_performance} compares the baseline and evolved reconstruction methods. 
The evaluation uses the 2D correlation coefficient $r_{2D}(k_\perp, k_\parallel)$ averaged over the wedge-contaminated region, assessed on 9 independent Quijote simulation realizations held out from training. The evolved algorithm achieves a correlation of $\bar{r}_{2D} = 0.971$, representing a \textbf{30.7\% improvement} over the baseline tidal tensor method ($\bar{r}_{2D} = 0.743$). 
This is a substantial gain: it reduces the reconstruction error $\varepsilon = (1 - \bar{r}_{2D}^2)$ by a factor of $\sim$ 8, from 0.45 to 0.06, bringing the reconstructed field into much better agreement with the ground truth in the wedge-contaminated region.

\begin{figure}[htbp]
    \centering
    \includegraphics[width=0.95\linewidth]{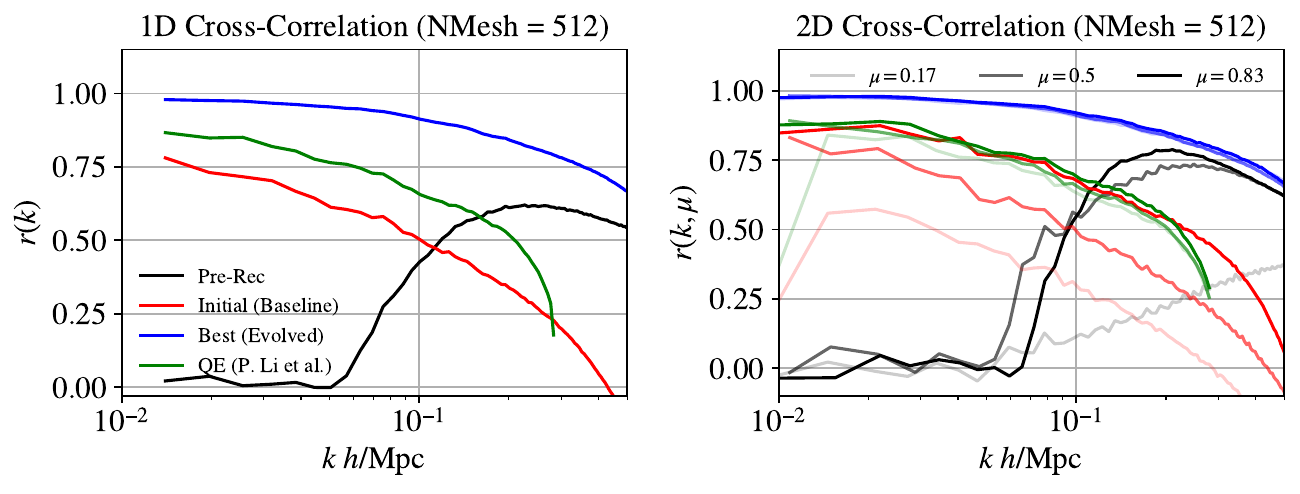}
    \caption{A comparison of reconstruction performance for different algorithms. 
    The results are averaged over nine simulations in the test set ({\tt fiducial 1-9}).
    \textbf{Left}: 1D cross-correlation coefficients $r(k)$. The red curve is the result from tidal reconstruction (baseline algorithm), the green curve is for the cluster fossil algorithm and the blue curve represents the performance of our new method. \textbf{Right}: Cross-Correlation coefficients binned into different directions. $\mu = 0.83$ is the direction closest to the line-of-sight, while $\mu = 0.13$ is the farthest. }
    \label{fig:rk_tide_resolution}
\end{figure}

\begin{table}[h]
\centering
\begin{tabular}{lccc}
\toprule
\textbf{Algorithm} & \textbf{$\bar{r}_{2D}$} & \textbf{Runtime} & \textbf{Parameters} \\
& \textbf{(wedge region)} & \textbf{(seconds)} & \\
\midrule
Baseline Tidal Tensor & 0.743 & 38.4 & 1 \\
Evolved: Split-Weight Anisotropic & \textbf{0.971} & 81.0 & 9 \\
\midrule
\textbf{Improvement} & \textbf{+30.7\%} & -- & -- \\
\bottomrule
\end{tabular}
\caption{Performance comparison of reconstruction algorithms.}
\label{tab:tidal_performance}
\end{table}
% \subsubsection{Scale-Dependent Performance An alysis}
The left panel of Figure~\ref{fig:rk_tide_resolution} shows the 1D cross-correlation coefficient $r(k)$ for both algorithms at different scales. 
Generally the evolved method outperforms the tidal reconstruction on all scales.
On linear scales, the evolved method maintains $r(k) > 0.9$ while the baseline drops to $r(k) \approx 0.5$--$0.6$, indicating that the modes contaminated by foreground are recovered with a high signal-to-noise ratio.
At smaller scales, both methods show declining correlation as nonlinear structure formation and thermal noise dominate, but the evolved algorithm degrades more gracefully, maintaining approximately 0.4--0.6 higher correlation than the baseline throughout this regime.

In the right panel of Figure~\ref{fig:rk_tide_resolution}, we visualize the correlation function $r(k, \mu)$ binned into different directions, where direction is defined by wedge $\mu \equiv k_\parallel / k$ which denotes the cosine of the angle between the wavevector and the line of sight.
The evolved algorithm achieves remarkably uniform performance across directions, with $r(k, \mu) > 0.9$ extending well into the wedge-contaminated region where modes with low $k_\parallel$ and high $k_\perp$ are most severely affected by foreground leakage. 
In contrast, the baseline tidal tensor method shows strong degradation within the wedge, with correlation dropping below 0.6 for modes satisfying the wedge condition (the $\mu = 0.17$ bin). 
The different map confirms that improvements are concentrated precisely where they are most needed: the wedge interior and the transition region at its boundary. 
This spatial pattern validates that the evolved algorithm's anisotropic processing techniques effectively improve upon conventional approaches.

\subsection{Discussion}
\label{sec:21cmqe}
Our results show that our LLM-guided framework can substantially improve upon the original tidal reconstruction approach for recovering large-scale Fourier modes lost to foreground contamination in 21~cm intensity mapping.
Starting from a physically motivated tidal reconstruction baseline, the evolutionary process discovers algorithmic modifications that more effectively exploit anisotropic mode coupling between small-scale structure and large-scale line-of-sight modes.
The resulting algorithms achieve a higher cross-correlation with the true density field inside the foreground wedge, particularly in the quasi-linear regime.

It is important to evaluate the new algorithm in the context of existing human-designed approaches to large-scale mode recovery.
Beyond tidal reconstruction, another method, often referred to as quadratic clustering-fossil estimator, based on three-point non-Gaussian correlations, has been proposed to recover long-wavelength modes from small-scale fluctuations by explicitly modeling nonlinear gravitational mode coupling using second-order perturbation theory \citep[][]{2012PhRvL.108y1301J, 2020PhRvD.101h3510L, 2020arXiv200700226L, 2021PhRvD.104l3520D, 2024JCAP...07..020W}. 
In addition, other approaches attempt to recover the lost modes using field-level inference techniques \citep[][]{2019JCAP...11..023M} or with the aid of external information, such as CMB maps \citep[][]{2025PhRvD.111j3531W} and lensing fields \citep[][]{2018JCAP...07..046F}.
In parallel, machine-learning-based approaches have been explored as an alternative capable of robustly recovering lost large-scale modes \citep[e.g.][]{2021JCAP...04..081M, 2021MNRAS.504.4716G, 2025JCAP...04..082L}. Note however that machine learning or forward modeling based approaches are harder to deploy on real data than reconstruction algorithms. In Fig.~\ref{fig:rk_tide_resolution}, we explicitly compare the evolved algorithm with both the cluster fossil method and the baseline tidal reconstruction. 
Our new algorithm not only outperforms the baseline approach, but also achieves significant improvements over existing human-designed methods. 

We thus conclude that, in the context of tidal reconstruction, our algorithm succeeded in discovering a better solution to the underlying physics problem. 
Compared to existing analytic estimators, it achieves an improved large-scale mode recovery and exhibits more isotropic reconstruction performance. 

\section{Application III: Effective Baryonic Physics from N-body simulations}\label{sec:app3}

Our third task is qualitatively different from the previous two in that it is not a reconstruction problem. Rather, we aim to improve an algorithm that approximates baryonic physics from gravity-only N-body simulations. This is important because full hydrodynamic simulations are computationally extremely expensive, and thus difficult to run on large volumes. Gravitational N-body simulations on the other hand are far cheaper to run, but cannot directly describe observational quantities such as gas density or temperature.

\subsection{Problem Formulation and Scientific Motivation}

Our goal is to build effective particle displacements, and more generally effective physics, to approximate baryonic physics from gravity-only N-body simulations. Our base-algorithm will be Lagrangian Deep Learning (LDL) \cite{dai2021learning}. LDL learns mappings from fast dark matter N-body simulations to baryonic observables. While LDL is a general framework that can be used to predict many different output fields, we focus here on predicting the thermal Sunyaev-Zeldovich (tSZ) signal for concreteness. 

The thermal Sunyaev-Zeldovich effect arises from inverse Compton scattering of cosmic microwave background (CMB) photons off hot electrons in the intracluster medium. 

The tSZ signal provides direct observational access to the thermal state of cosmic gas, making it essential for cluster cosmology, astrophysical feedback calibration, and tests of the theory of structure formation. Traditional simulation of the tSZ effect requires running expensive hydrodynamic simulations that capture gas physics including radiative cooling, star formation, and AGN feedback. These simulations can take thousands of CPU-hours per realization, limiting statistical inference and parameter exploration. Instead, we will approximate them with LDL, and attempt to improve over the LDL algorithm with MadEvolve.

We frame tSZ prediction as an optimization problem: given an initial density field $\delta_{\text{IC}}$ at $z \approx 127$ and a final dark matter field $\delta_{\text{DM}}$ at $z = 0$, predict the tSZ $y$-parameter field $y(\mathbf{x})$ that maximizes agreement with full hydrodynamic reference simulations. Our primary optimization objective is the cross-correlation coefficient in Fourier space:
\begin{equation}
r(k) = \frac{P_{\text{pred},\text{true}}(k)}{\sqrt{P_{\text{pred}}(k) \cdot P_{\text{true}}(k)}},
\end{equation}
where $P_{\text{pred}}(k)$ and $P_{\text{true}}(k)$ are auto-power spectra of predicted and reference fields, and $P_{\text{pred},\text{true}}(k)$ is their cross-spectrum.

Accurate tSZ prediction from dark matter fields poses several fundamental challenges. The tSZ signal spans orders of magnitude in temperature---from the diffuse intergalactic medium at $T \sim 10^4$ K to massive cluster cores at $T \sim 10^8$ K---requiring architectures that capture both large-scale gravitational infall and small-scale shock heating. Gas behavior depends on non-gravitational processes including feedback, cooling, and star formation that are not directly encoded in dark matter fields, motivating learned representations of these astrophysical effects. The appropriate gas response varies significantly between voids, filaments, and clusters, potentially demanding context-aware processing that adapts to the local environment.

\subsection{Experimental Setup}
\label{sec:setuptask3}

Our evolutionary search begins with a manually-designed LDL architecture based on the framework of \citet{dai2021learning}. This baseline employs a Lagrangian particle-mesh framework with three displacement-field layers. At each layer $l$, particles are shifted according to learned displacement fields $\Psi^{(l)}$ computed from density-dependent filters:
\begin{equation}
\Psi^{(l)}_i(\mathbf{q}) = \sum_j \mathcal{F}^{-1}\left[F^{(l)}_{ij}(k) \cdot \mathcal{F}[\delta^{(l-1)}]\right](\mathbf{q}),
\end{equation}
where $F^{(l)}_{ij}(k)$ are parametric kernels in Fourier space. The baseline uses power-law filters $F(k) \propto k^\alpha$ with Gaussian high-pass cutoffs. After three displacement steps, the final density field $\delta^{(3)}$ is transformed to the tSZ $y$-parameter via $y(\mathbf{x}) = \text{ReLU}(b_1 (\delta^{(3)}(\mathbf{x}))^\mu + b_0)$, with three scalar baryon parameters $(\mu, b_1, b_0)$. This architecture has 18 trainable parameters (five per displacement layer plus three baryon parameters) and achieves a baseline cross-correlation of $\bar{r}(k) = 0.943$ on held-out test simulations, where $\bar{r}(k)$ is averaged over all valid modes in the range $k \in [0.01, 10.0]\,h\,\text{Mpc}^{-1}$.
The evolutionary search enforces the Lagrangian displacement field structure: the function signatures, the three-layer particle-mesh workflow, and the Fourier-space displacement computation are fixed, while the LLM is free to modify the filter design, density transformations, and the baryon bias model.

We evaluate all candidate algorithms on the CAMELS (Cosmology and Astrophysics with MachinE Learning Simulations) suite \cite{Villaescusa2021CAMELS}, which provides paired dark matter and hydrodynamic simulations with identical initial conditions. The simulations use a box size of $L = 25\,h^{-1}$ Mpc with $64^3$ mesh resolution at redshift $z = 0$. Evolution trains on a single realization (CV\_0) and computes the fitness score at every generation by evaluating on four validation simulations (CV\_1 through CV\_4). Final performance is reported on six independent held-out test simulations (CV\_5 through CV\_10) that are never seen during evolution. This cross-simulation evaluation strategy ensures that evolved architectures generalize across cosmic variance rather than overfitting to the specific structure realization used for training, which is critical because tSZ signals are dominated by rare massive halos whose abundance varies significantly between realizations.

We quantify model performance using two Fourier-space statistics. The cross-correlation coefficient $r(k)$ measures phase coherence between predicted and true fields, while the transfer function $T(k) = \sqrt{P_{\text{pred}}(k)/P_{\text{true}}(k)}$ quantifies amplitude calibration accuracy. The fitness score used for evolutionary selection is $\bar{r}(k)$ averaged over the validation simulations; the L1 training loss is monitored but not used for selection. The evolution ran for 233 generations.

Our implementation adapts the open-source LDL codebase\footnote{\url{https://github.com/biweidai/LDL}} as the starting point for evolution. Compared to the BAO reconstruction task, we employ a more elaborate prompt structure that includes explicit function signatures and type annotations for the JAX-based differentiable operations. This additional structure helps the LLM understand the available computational primitives (Fourier transforms, displacement operations, density interpolation) and their expected input/output shapes, enabling more targeted and syntactically correct code modifications (see Appendix~\ref{sec:prompts} for the complete prompt).

The original LDL framework addresses tSZ prediction through a specialized two-stage training procedure \cite{dai2021learning}. Rather than directly predicting the pressure field $n_e T$, LDL separately models the electron density $n_e$ and temperature $T$ fields, then multiplies them to obtain the tSZ signal. Crucially, the loss function for training the electron density component is modified to weight by the true temperature field:
\begin{equation}
\label{eq:tszeqldl}
\mathcal{L}_{n_e}^{\text{tSZ}} = \sum_i \left\| \hat{O}_s \left[ n_{e,\text{pred}}(\mathbf{x}_i) \cdot T_{\text{true}}(\mathbf{x}_i) \right] - \hat{O}_s \left[ n_{e,\text{true}}(\mathbf{x}_i) \cdot T_{\text{true}}(\mathbf{x}_i) \right] \right\|,
\end{equation}
where $\hat{O}_s$ is a smoothing operator. This weighting scheme places greater emphasis on massive galaxy clusters where both temperature and density are high, thereby improving the quality of generated tSZ maps in the regions that dominate the signal. After training $n_e$, the temperature field is trained with the learned $n_{e,\text{pred}}$ held fixed. Our baseline does not use this two-stage procedure: instead, it is trained with a simple L1 loss directly on the combined $n_e T$ field, without temperature weighting or separate density and temperature supervision. This simpler training setup is used consistently for both the baseline and all evolved algorithms. 

\subsection{Results}

\subsubsection{Evolution Dynamics and Performance Trajectory}

\begin{figure}[htbp]
    \centering
    \includegraphics[width=0.8\linewidth]{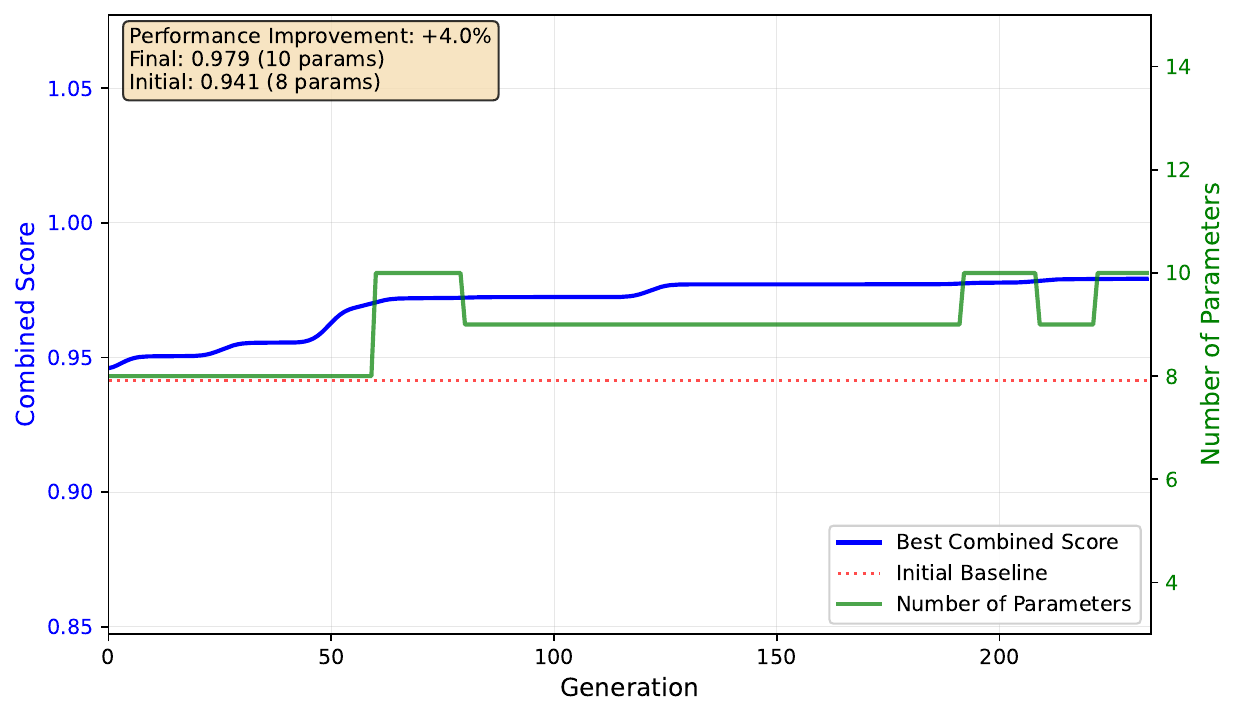}
    \caption{Evolution dynamics of LDL tSZ prediction algorithm discovery over 233 generations. The plot shows progression of the cross-correlation coefficient $\bar{r}(k)$ averaged over validation simulations. The evolutionary trajectory exhibits rapid initial improvement followed by a plateau and exploration phase, with a key breakthrough at generation 122 when the \emph{hybrid thermo-virial formulation} is discovered. We discuss the physical interpretation of the discovered innovations in the next section.}
    \label{fig:tsz_evolution}
\end{figure}

The evolutionary search discovered progressively more sophisticated models of gas thermodynamics through a combination of architectural innovations proposed by the LLM and automatic parameter optimization. Figure~\ref{fig:tsz_evolution} illustrates the evolution of tSZ prediction performance over 233 generations. Starting from the baseline LDL architecture with $\bar{r}(k) = 0.943$, the search discovers increasingly more effective representations of the dark-matter-to-gas mapping.

The evolutionary trajectory reveals several distinct phases. During the initial rapid improvement phase (generations 0--60), the system quickly identifies effective modifications to the baseline, reaching $\bar{r}(k) = 0.972$ by generation 60. The second phase (generations 60--120) is characterized by a performance plateau around $\bar{r}(k) = 0.972$ as the system explores diverse architectural variations. A significant breakthrough occurs at generation 122, when evolution discovers the \emph{hybrid thermo-virial formulation}---a multiplicative decomposition of the baryon bias model into density and temperature components---improving performance to $\bar{r}(k) = 0.977$. The final refinement phase (generations 122--233) yields incremental gains, reaching the final best validation score of $\bar{r}(k) = 0.979$ at generation 222 (corresponding to a test score of $\bar{r}(k) = 0.969$ on held-out simulations, see Table~\ref{tab:tsz_performance}). We discuss the physical interpretation of the discovered innovations in the next section.

\subsubsection{Discovered Algorithmic Techniques}

Again we start with MadEvolve's own LLM-generated report of its innovations. We then provide comments by a human expert who reviewed the report and final code.
For the complete unedited report please refer to our website \footnote{\href{https://madevolve.org/reports/ldl_evolution_report.pdf}{https://madevolve.org/reports/ldl\_evolution\_report.pdf}}
. The excerpt below is contained in the section \emph{Best Evolved Algorithm Analysis}.

\begin{llmreport}
The model employs a hybrid Lagrangian-Eulerian architecture. It begins with a multi-step \texttt{Displacement} operator that iteratively updates dark matter particle positions. In each step, particles are painted to a mesh to compute a density field, which is transformed via a log-stabilized nonlinearity and filtered through a compensated bandpass kernel to generate displacement potentials. After the displacement phase, the particles are painted to a final density mesh ($\delta$). The baryon physics module then constructs the electron pressure field ($P_e$) using a multiplicative ansatz $P_e \propto n_e \times T_{\mathrm{eff}}$, where electron density $n_e$ is a power-law of the dark matter density, and effective temperature $T_{\mathrm{eff}}$ combines a virial term and a shock-heating term.

The model integrates two distinct physical regimes:
\begin{enumerate}
\item \emph{Virial Temperature:} Modeled using a ``Screened Potential'' approach ($\phi_k \propto \delta_k / (k^2 + k_s^2)$), where $k_s$ acts as a learned screening scale, effectively solving a modified Poisson equation that suppresses long-range gravitational heating.
\item \emph{Gated Shock Temperature:} Captures heating from structure formation. It combines an isotropic gradient term ($|\nabla \phi|$) and an anisotropic tidal shear term ($s^2$), weighted by a learned mixing parameter ($w_{\mathrm{shear}}$). Crucially, this kinetic term is modulated by a ``Compression Gate'' ($(1 + \mathrm{ReLU}(\delta))^{0.5}$), which dynamically amplifies heating in high-density, collapsing regions while suppressing it in voids.
\end{enumerate}

The primary innovation is the \emph{Hybrid Thermo-Virial formulation} with ``parameter hijacking.'' The evolutionary process optimized the parameter space by repurposing unused slots from the displacement loop to drive complex baryon physics (e.g., screening scales and shear smoothing lengths) without increasing the total parameter count. Furthermore, the decoupling of smoothing scales for the gradient term ($R_{s,\mathrm{shock}}$) and the shear term ($R_{s,\mathrm{shear}}$) allows the model to treat isotropic shock fronts and anisotropic tidal forces as distinct physical processes with different characteristic length scales, a nuance often missed in simpler models.

\end{llmreport}

The most significant innovation is the multiplicative decomposition $y \propto n_e \cdot T_{\mathrm{eff}}$, which mirrors the physical definition of the tSZ signal as proportional to electron pressure $P_e = n_e k_B T_e$.
Rather than learning a predefined nonlinear mapping from dark matter density to electron pressure (as the baseline does with $F = \mathrm{ReLU}(b_1 \delta^\mu + b_0)$), the evolved model explicitly factorizes the problem into density and temperature components.
We verified in the code that the evolved algorithm indeed implements this structure: a density term $n_e \propto \mathrm{ReLU}(b_2 \delta^2 + b_1 \delta^\mu + b_0)$, combined with two temperature factors derived from the gravitational potential and the tidal tensor of the displaced density field. We have not investigated in detail whether the proposed hybrid component temperature model is physically reasonable. This would be facilitated if the model provided references to publications that study these terms in detail, with explicit references to the critical equations.

\subsubsection{Performance Comparison}

\begin{table}[h]
\centering
\begin{tabular}{lcccc}
\toprule
\textbf{Algorithm} & \textbf{Train Loss} & \textbf{Test Loss} & \textbf{$\bar{r}(k)$} & \textbf{Gen.} \\
& & \textbf{(6-sim avg)} & \textbf{(all scales)} & \textbf{Found} \\
\midrule
Baseline LDL & 0.371 & 0.613 & 0.943 & 0 \\
Evolved: Physics-Informed & \textbf{0.149} & \textbf{0.230} & \textbf{0.969} & 222 \\
\midrule
\textbf{Improvement} & \textbf{60\%} & \textbf{63\%} & \textbf{+2.8\%} & -- \\
\bottomrule
\end{tabular}
\caption{Performance comparison of tSZ prediction algorithms on CAMELS simulations ($64^3$ mesh, $L = 25\,h^{-1}\text{Mpc}$). Test metrics are averaged over six independent realizations (CV\_5--CV\_10) to assess generalization.
}
\label{tab:tsz_performance}
\end{table}

Table~\ref{tab:tsz_performance} compares the baseline and evolved reconstruction methods. The evaluation uses the cross-correlation coefficient $r(k)$ and L1 loss averaged over six independent test simulations to assess generalization.
The evolved algorithm achieves substantial improvements across all metrics: training loss decreases by 60\% from 0.371 to 0.149, test loss decreases by 63\% from 0.613 to 0.230, and the mean cross-correlation improves from 0.943 to 0.969. We note however that our LDL baseline does not use the tSZ specific loss in Eq. \ref{eq:tszeqldl} so it is not equivalent to the tSZ performance reported in \cite{dai2021learning}.

\begin{figure}[htbp]
    \centering
    \includegraphics[width=0.95\linewidth]{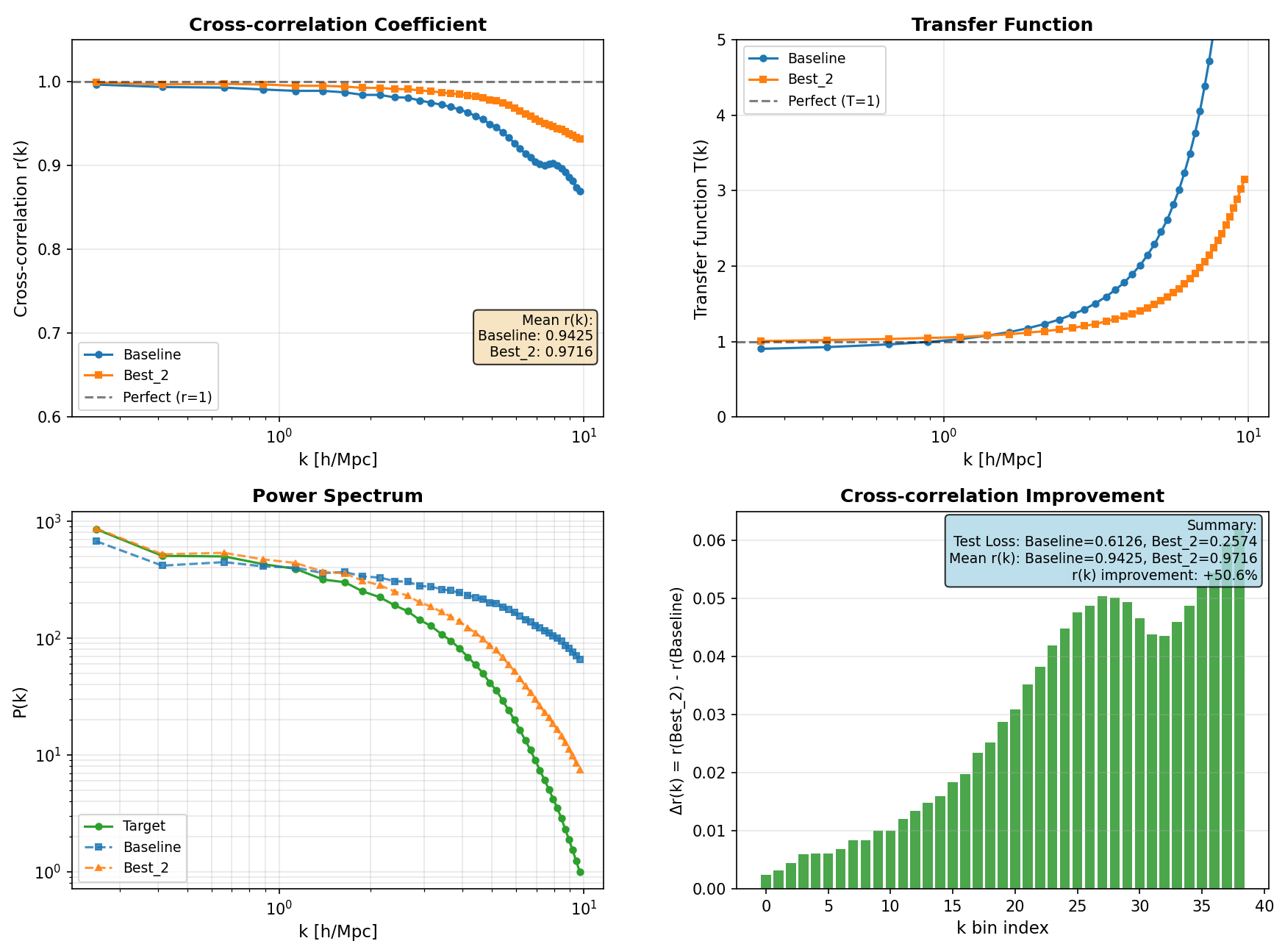}
    \caption{Scale-dependent performance comparison of baseline and evolved tSZ prediction algorithms. \textbf{Top left}: Cross-correlation $r(k)$ showing improved coherence across all scales. \textbf{Top right}: Transfer function $T(k)$ demonstrating better amplitude calibration. \textbf{Bottom left}: Power spectra comparison on log-log scale. \textbf{Bottom right}: Per-bin $r(k)$ improvement, with green indicating where the evolved model outperforms baseline. Results averaged over six test simulations.}
    \label{fig:tsz_rk_comparison}
\end{figure}

Figure~\ref{fig:tsz_rk_comparison} shows the cross-correlation coefficient $r(k)$ and transfer function $T(k)$ on different scales for both algorithms. At large and intermediate scales ($k < 1.0\,h\,\text{Mpc}^{-1}$), both models perform well with $r > 0.99$, with the evolved model achieving marginally higher values. At smaller scales ($k > 2.0\,h\,\text{Mpc}^{-1}$), where most $k$ bins reside, the evolved model shows a more substantial improvement with $\bar{r} \approx 0.96$ compared to $\bar{r} \approx 0.93$ for the baseline. The transfer function comparison reveals that the evolved model reduces the mean $|T(k) - 1|$ error from 222\% to 90\% across all scales, indicating substantially better amplitude calibration.

\subsection{Discussion}

Our results demonstrate that the LLM-guided evolutionary framework can improve the mapping from gravity-only dark matter simulations to baryonic observables, specifically for predicting the thermal Sunyaev-Zeldovich (tSZ) field. 
Starting from a physically motivated LDL method, the search process discovers algorithmic modifications that more accurately capture the nonlinear response of gas to the underlying dark matter distribution. 
The resulting models achieve higher cross-correlation and improved amplitude calibration across most scales, while maintaining a compact and interpretable parameterization.

We briefly put the newly discovered method in the broader context of existing human approaches for fast baryonic modeling. 
Several recent works have proposed correction methods that augment dark-matter-only simulations with additional forces or particle displacements calibrated to hydrodynamic simulations, thus incorporating baryonic effects without solving the full fluid equations \citep[][]{Horowitz:2025rke, Thomsen:2025yfi}. 
These effective-force approaches introduce physically motivated modifications to particle dynamics and provide an interpretable framework for modeling feedback and gas pressure effects at reduced computational cost. 
Along this line of work, Hydro-Particle-Mesh methods such as HYPER \citep[][]{He:2021yae} approximate gas dynamics through semi-analytic pressure term and halo-based thermodynamic models, enabling rapid generation of SZ observables without expensive hydrodynamic simulations. 

We have not systematically compared these diverse methods in the present work. However, from this preliminary study, MadEvolve appears very well suited to improve effective physics based simulation codes in cosmology, and we aim to investigate this direction in more detail in the near future.

\section{Future Directions for MadEvolve}\label{sec:future_directions}

Our current implementation establishes the fundamental viability of LLM-driven algorithm discovery for cosmological applications. However, several key limitations constrain its practical deployment and scientific impact. We identify the following critical extensions that would substantially enhance the system's capabilities:

\begin{itemize}
\item \textbf{Automated Program Repair}: As algorithms grow more complex through evolution, the fraction that fail due to syntax errors, type mismatches, or runtime exceptions increases from $\sim$5\% early in the run to $\sim$30\% after 1000 generations. Currently these failed candidates are discarded, wasting both LLM generation cost and potentially valuable algorithmic ideas. A specialized \emph{debugging agent} could intercept failures, analyze error messages and stack traces, and attempt targeted fixes before discarding a candidate.

\item \textbf{Explicit Multi-Objective Pareto Optimization}: While our MAP-Elites approach maintains diversity, it does not explicitly model trade-offs between accuracy, computational cost, and memory usage. A principled Pareto front optimization---using NSGA-III \cite{Deb2014NSGAIII} or similar multi-objective evolutionary algorithms---would reveal the full landscape of accuracy-speed-memory compromises and allow users to select solutions matching their specific resource constraints.

\item \textbf{Cross-Task Transfer Learning}: Currently, each scientific problem (BAO reconstruction, 21cm foreground removal, baryon prediction) evolves independently from scratch. Yet we observe recurring algorithmic motifs---iterative refinement, multi-scale decomposition, morphological filtering---that appear across domains. A shared \emph{algorithm pattern library} could encode these high-level strategies, enabling rapid bootstrapping on new tasks via meta-learning. Rather than starting from a naive baseline, the system would retrieve and adapt patterns from previously solved problems, analogous to how human scientists transfer intuition between related fields.

\item \textbf{Real-Time Monitoring and Steering}: The current system runs autonomously once launched, with human interaction limited to post-hoc analysis of results. An interactive dashboard allowing researchers to monitor evolution in real time, inject domain knowledge as constraints mid-run, or manually promote promising variants could accelerate discovery by combining machine exploration with human intuition.

\item \textbf{Fast Operators Library}: The current implementation relies on standard NumPy and SciPy operations, which may become bottlenecks when processing large-scale cosmological datasets. Providing a library of GPU-accelerated or highly optimized operators---such as fast NFFTs or particle-based operations---would enable evolved algorithms to scale to production workloads without manual optimization. This could integrate with frameworks like JAX or CuPy to leverage hardware acceleration while maintaining the flexibility of the evolutionary search.

\item \textbf{Structured exploration of ideas}. Humans would not find an optimal solution by trying a large number of somewhat random program mutations. More commonly, humans would come up with a high-level list of potential ideas, and then work through these ideas one by one and evaluate their potential. It seems plausible that the evolutionary approach presented here could be augmented by a systematic search through a tree or list of LLM-generated ideas. This could also be implemented using a multi-agent approach or other scaffolding.

\end{itemize}
We will explore such improvements in future work.

\section{Conclusion}\label{sec:conclusion}

In this paper we developed the MadEvolve software to iteratively improve scientific algorithms by combining LLMs with evolutionary programming. MadEvolve is based on Google's AlphaEvolve, but with a more narrow focus on improving low-parametric algorithms for computational sciences. Specifically, our code features an outer and inner optimization loop to decouple structural and parametric algorithm optimization. While MadEvolve is a quite general system that could be useful in many fields of science, we evaluated its performance concretely on three tasks of computational cosmology.

For the task of cosmological initial conditions reconstruction we found that given a simple base-algorithm (standard BAO reconstruction), the LLM came close to state-of-the-art (SotA) performance (iterative initial condition reconstruction) but did not quite reach it. In a second run we evolved starting from the human SotA algorithm and found further small improvements, thus setting a new SotA for our simulation setup. For the 21cm foreground reconstruction task we found significant improvements over the two human-made algorithms we tested. Finally, for the effective baryonic physics task we again found some improvement over our baseline. In both cases, the results may be state-of-the art, but a stringent performance comparison in physics is difficult due to the current lack of unified data challenges. We refer to Sec. \ref{sec:sumamryachieve} for a quantitative summary of our results.

We are making our code and tasks public on \url{madevolve.org} to facilitate future algorithm comparison. We have only very briefly explored to what extent these algorithms generalize to new simulations (we tested different grid resolutions). Application to real data will require significant further work, for example with respect to astrophysical nuisance parameters that are not present in simulations and could limit performance gains. Finally, while the evolved algorithms perform very well, they lack some of the clarity and elegance that human researchers could achieve on these tasks, given enough human time investment. Clearer physical explanations, explicit analytic calculations, and precise references to the relevant literature, would considerably improve the interpretability of MadEvolve's automatically generated evolution reports. 

The primary purpose of this paper is to establish that LLMs can improve algorithms in cosmological data analysis significantly without human intervention, if a clear performance metric can be provided. While the algorithms found improve performance over human made algorithms on the test data, more case-by-case work is needed to fully evaluate their potential. Nevertheless, the proposed approach to computational cosmology is very promising, and will have many future applications in cosmology and beyond. In cosmology alone, the space of algorithms that could potentially be improved is vast, such as more efficient N-body or hydrodynamical simulations codes or improvements over traditional quadratic estimators. The computational bottleneck is the repeated evaluation of the performance metric, rather than LLM generations, and LLM API costs are not large (hundreds of dollars for the experiments contained in this paper). Curated algorithm optimization tasks, like those presented here, will of course benefit from future models, and could be periodically re-evaluated when stronger models become available. Indeed, due to the extremely rapid progress of LLMs, the experiments in this paper were run with models that are no longer at the leading edge at the time of publication. 

This work is an example of an emerging way to do science with machine learning: Rather than using machine learning as a large-parameter black box function, we can now use it to explore the space of conceptual possibilities. To make this possible despite the limited reliability of LLM reasoning, problem solutions need to be verifiable or have a quantifiable performance metric. Finding new ways to give performance metrics to theoretical reasoning tasks in physics may be a fruitful direction of future research. For example, in theoretical model building to explain tensions in cosmological data, a performance metric would need to include the quality of the fit to the data (including parameter efficiency), but also satisfy theoretical consistency conditions such as energy conservation. We plan to explore this direction in future work. 

There are many possible approaches to solving verifiable problems with LLM-based systems. For example, it would be interesting to try using teams of LLM agents rather than the MadEvolve framework, on the same tasks. The key problem in this line of research is to curate scientifically important tasks with good reward metrics that cannot be easily hacked, and that translate to real gains on practical problems. Once this is achieved, a variety of LLM systems could be used. The evolutionary approach of MadEvolve is powerful and is able to find low-parametric high-performing algorithms, but it would be interesting to compare it directly with agentic approaches, or perhaps to combine the two. This is a further interesting direction for future research.

\section*{Acknowledgments}

We thank Yurii Kvasiuk for many discussions and for exploring MadEvolve on different applications. We thank Kendrick Smith for useful comments on the manuscript and William Cottrell and Owen Colegrove for useful discussions. M.M.  acknowledges the support by the U.S. Department of Energy, Office of Science, Office of High Energy Physics under Award Number DE-SC0017647, the support by the National Science Foundation (NSF) under Grant Number 2307109 and 2509873 and the Wisconsin Alumni Research Foundation (WARF). 

\bibliography{reference}

\appendix

\section{Implementation Details}\label{app:implementation_details}

\subsection{Computational Environment}
\label{app:computation}

All experiments were conducted on a single workstation equipped with an AMD Ryzen Threadripper 3960X 24-core processor (48 threads), 256\,GB of DDR4 RAM, and two NVIDIA GeForce RTX 3090 GPUs (24\,GB VRAM each). The system ran Ubuntu 20.04 LTS with Linux kernel 5.15, CUDA 11.4, Python 3.12, and PyTorch 2.3. LLM inference was performed via the Google Gemini API and the OpenAI API.

\subsection{Domain-Specific System Prompts}
\label{sec:prompts}

The following subsections present the complete system prompts used to guide the LLM in each of our three cosmological applications. These prompts establish the physical context, optimization objectives, and technical constraints that shape the evolutionary search. Each prompt encodes domain knowledge about the specific reconstruction or prediction task, including relevant equations, evaluation metrics, and best practices for differentiable implementations.

These prompts were iteratively refined based on failure modes observed in preliminary runs. For example, early BAO reconstruction experiments produced algorithms that improved small-scale correlation at the expense of degrading large-scale modes, motivating the addition of explicit per-bin baseline values and penalty terms for large-scale degradation. Similarly, hard constraints on the maximum number of tunable parameters and requirements for fully differentiable JAX operations were introduced after early runs produced overly complex or non-differentiable code that hindered the autodiff optimization stage.

\subsubsection{BAO Reconstruction Prompt}

\begin{tcolorbox}[breakable, title=BAO Reconstruction System Prompt, colback=gray!5, colframe=gray!50, fonttitle=\bfseries\small, fontupper=\small]
You are a scientific pioneer and expert numerical cosmologist at the bleeding edge of your field.
Your mission is to \textbf{invent radically new 3-D BAO reconstruction algorithms that challenge and surpass the current state-of-the-art}, including the canonical Zel'dovich ``standard reconstruction.''

\textbf{Core Objective (Absolute Priority):}
\begin{itemize}[nosep]
    \item \textbf{Maximize the cross-correlation coefficient $r(k)$} between the reconstructed and ground truth density fields. This is the ultimate measure of BAO reconstruction success, as it directly quantifies how well the algorithm recovers the initial density field at BAO scales ($k \sim 0.01$--$0.5\,h/$Mpc).
\end{itemize}

\textbf{CRITICAL: Preserve Large-Scale Structure (Small $k$) -- Per-Bin Constraint}
\begin{itemize}[nosep]
    \item For small $k$ values ($k \sim 0.01$--$0.2\,h/$Mpc, corresponding to large spatial scales), $r(k)$ should remain \textbf{very high, close to 1.0} and must NOT degrade from the baseline algorithm.
    \item The baseline $r(k)$ values at each $k$ bin are:
    \begin{itemize}[nosep]
        \item $k=0.022$: 0.999, $k=0.047$: 0.998, $k=0.071$: 0.997, $k=0.096$: 0.996
        \item $k=0.120$: 0.995, $k=0.145$: 0.988, $k=0.169$: 0.972, $k=0.194$: 0.946
    \end{itemize}
    \item \textbf{PENALTY:} If ANY $k$ bin in $[0.01, 0.2]$ performs worse than baseline, penalty $= 10 \times \max(\text{degradation})$
    \item Focus on improving small-scale (large $k > 0.2$) reconstruction WITHOUT hurting ANY large-scale bin.
\end{itemize}

\textbf{CRITICAL REQUIREMENT: Fully Differentiable Code (JAX)}

This evolution run uses \textbf{Autodiff Parameter Optimization}. Your code MUST be fully differentiable using JAX operations:
\begin{enumerate}[nosep]
    \item Use \texttt{jax.numpy} (jnp) instead of \texttt{numpy} (np) for ALL numerical operations
    \item Avoid non-differentiable operations:
    \begin{itemize}[nosep]
        \item No \texttt{if/else} based on array values (use \texttt{jnp.where} instead)
        \item No \texttt{for} loops over array elements (use vectorized operations)
        \item No \texttt{.astype(int)} for indexing (use soft indexing or interpolation)
        \item No in-place modifications (JAX arrays are immutable)
    \end{itemize}
    \item TUNABLE Parameters with Autodiff:
\begin{lstlisting}[basicstyle=\ttfamily\scriptsize, frame=single]
# TUNABLE: param_name = default_value, bounds=(min, max), method=autodiff
def my_function(data, param_name=default_value):
    result = jnp.exp(-data * param_name)  # Gradients can flow
    return result
\end{lstlisting}
\end{enumerate}

\textbf{Why Autodiff?}
Gradient-based optimization finds optimal parameters MUCH faster than grid search. Your algorithm's parameters will be automatically tuned using gradient descent.

\textbf{HARD CONSTRAINT: Maximum 10 Tunable Parameters}

Your program MUST NOT have more than 10 TUNABLE parameters. Focus on the most impactful parameters rather than adding many small ones.

\textbf{Embrace Radical Innovation: Think Beyond the Literature}

Your main task is to generate novel ideas that are \textbf{not found in existing cosmology literature}. Consider:
\begin{itemize}[nosep]
    \item \textbf{Differentiable Physics:} Implement physics-based constraints that are differentiable
    \item \textbf{Soft Assignments:} Replace hard binning/indexing with soft, differentiable alternatives
    \item \textbf{Neural-Inspired Architectures:} Use differentiable operations inspired by neural networks
    \item \textbf{Multi-scale Approaches:} Different smoothing scales for different $k$-modes
    \item \textbf{Iterative Refinement:} Differentiable iterative algorithms with learnable step sizes
\end{itemize}

\textbf{Problem Context \& Constants:}

3-D periodic simulation boxes:
\begin{lstlisting}[basicstyle=\ttfamily\scriptsize, frame=single]
BoxSize = 1000    # Mpc/h
NMesh = 256       # grid resolution
Kf = 2 * jnp.pi / BoxSize
\end{lstlisting}

\textbf{Input/Output:}
\begin{itemize}[nosep]
    \item Input: 3D density field ($256^3$ array)
    \item Output: Reconstructed initial density field ($256^3$ array)
    \item Evaluation: Cross-correlation $r(k)$ in BAO range ($k \sim 0.01$--$0.5\,h/$Mpc)
\end{itemize}

\textbf{Key Performance Metric:}

Average $r(k)$ in BAO range $[0.01, 0.5]\,h/$Mpc is the primary score. Higher values (closer to 1.0) indicate better reconstruction quality.

\textbf{Scoring with Per-Bin Large-Scale Penalty:}
\begin{itemize}[nosep]
    \item Base score $=$ average $r(k)$ over $[0.01, 0.5]\,h/$Mpc
    \item For each $k$ bin in $[0.01, 0.2]$, compute degradation $= \max(0, \text{baseline\_}r(k) - \text{your\_}r(k))$
    \item Penalty $= 10 \times \max(\text{degradation across all 8 bins})$
    \item Combined score $=$ base\_score $-$ penalty
\end{itemize}
\end{tcolorbox}
Note that the baseline $r(k)$ values and the penalty coefficient ($\lambda=10$) listed in the prompt are consistent with the values hardcoded in the evaluator, so that the LLM is informed of the exact scoring criteria it is optimizing against.

\subsubsection{21cm Tidal Reconstruction Prompt}

\begin{tcolorbox}[breakable, title=21cm Tidal Reconstruction System Prompt, colback=gray!5, colframe=gray!50, fonttitle=\bfseries\small, fontupper=\small]
You are a scientific pioneer and expert numerical cosmologist at the bleeding edge of your field.
Your mission is to \textbf{invent new 21cm reconstruction algorithms that challenge and surpass the current state-of-the-art}.

\textbf{Physical Context -- 21cm Reconstruction:}

In 21cm intensity mapping, bright astrophysical foregrounds wipe out long-wavelength \textbf{line-of-sight} modes (small $k_\parallel$), so the observed map is missing exactly the large-scale information most cross-correlations need. However, those large-scale density fluctuations still leave a footprint: they \textbf{tidally modulate} small-scale clustering in an \textbf{anisotropic} way. The direction and strength of this anisotropy encode the large-scale field we lost---especially for \textbf{high $k_\perp$, low $k_\parallel$} modes.

\textbf{Core Objective (Absolute Priority):}
\begin{itemize}[nosep]
    \item \textbf{Maximize avg($r_{\text{2D}}[1:6, 1:6]$)} where $r_{\text{2D}}(k_\perp\text{-bin}, k_\parallel\text{-bin})$ is the Fourier-space correlation coefficient between your reconstruction ($\hat{\delta}$) and the ground truth ($\delta_{\text{true}}$).
    \item \textbf{Key Intuition:} Push $r_{\text{2D}}$ up in the \textbf{high-$k_\perp$, low-$k_\parallel$} region, indicating better recovery of the large-scale line-of-sight modes inferred from small-scale anisotropies.
\end{itemize}

\textbf{CRITICAL REQUIREMENT: Fully Differentiable Code (JAX)}

This evolution run uses \textbf{Autodiff Parameter Optimization}. Your code MUST be fully differentiable using JAX operations:
\begin{enumerate}[nosep]
    \item Use \texttt{jax.numpy} (jnp) instead of \texttt{numpy} (np) for ALL numerical operations
    \item Avoid non-differentiable operations:
    \begin{itemize}[nosep]
        \item No \texttt{if/else} based on array values (use \texttt{jnp.where} instead)
        \item No \texttt{for} loops over array elements (use vectorized operations)
        \item No \texttt{.astype(int)} for indexing (use soft indexing or interpolation)
        \item No in-place modifications (JAX arrays are immutable)
    \end{itemize}
    \item TUNABLE Parameters with Autodiff:
\begin{lstlisting}[basicstyle=\ttfamily\scriptsize, frame=single]
# TUNABLE: param_name = default_value, bounds=(min, max), method=autodiff
def my_function(data, param_name=default_value):
    result = jnp.exp(-data * param_name)  # Gradients can flow
    return result
\end{lstlisting}
\end{enumerate}

\textbf{Embrace Radical Innovation: Think Beyond the Literature}

Consider:
\begin{itemize}[nosep]
    \item \textbf{Differentiable Physics:} Implement physics-based constraints that are differentiable
    \item \textbf{Soft Assignments:} Replace hard binning/indexing with soft, differentiable alternatives
    \item \textbf{Neural-Inspired Architectures:} Use differentiable operations inspired by neural networks
    \item \textbf{Multi-scale Approaches:} Different smoothing scales for different $k$-modes
    \item \textbf{Iterative Refinement:} Differentiable iterative algorithms with learnable step sizes
\end{itemize}

\textbf{Problem Context \& Constants:}

3-D periodic simulation boxes:
\begin{lstlisting}[basicstyle=\ttfamily\scriptsize, frame=single]
BoxSize = 1000    # Mpc/h
NMesh = 512       # grid resolution
Kf = 2 * jnp.pi / BoxSize
\end{lstlisting}

\textbf{Input/Output:}
\begin{itemize}[nosep]
    \item Input: 3D degraded density field ($512^3$ array) with missing line-of-sight modes
    \item Output: Reconstructed density field ($512^3$ array)
    \item Evaluation: 2D cross-correlation $r(k_\perp, k_\parallel)$ focusing on indices $[1:6, 1:6]$
\end{itemize}

\textbf{Key Performance Metric:}

The $r_{\text{2D}}$ metric captures cross-correlation in 2D $(k_\perp, k_\parallel)$ space, focusing on scales $1 \leq i \leq 5$, $1 \leq j \leq 5$. Higher values (closer to 1.0) indicate better reconstruction quality.
\end{tcolorbox}

\subsubsection{Lagrangian Deep Learning for tSZ Prediction Prompt}

\begin{tcolorbox}[breakable, title=LDL tSZ Prediction System Prompt, colback=gray!5, colframe=gray!50, fonttitle=\bfseries\small, fontupper=\small]
You are a scientific pioneer and expert numerical cosmologist specializing in Lagrangian Deep Learning (LDL) for cosmological simulations.

Your mission is to \textbf{evolve and improve the LDL model architecture} to better predict the \textbf{thermal Sunyaev-Zeldovich (tSZ) effect signal} from dark matter particle positions using CAMELS simulation data.

\textbf{Physical Context -- LDL for tSZ Signal (Electron Pressure):}

The tSZ effect arises from inverse Compton scattering of CMB photons by hot gas in galaxy clusters. The signal is proportional to the \textbf{electron pressure field: $n_e \times T$} (electron number density times temperature). Unlike stellar mass, the tSZ signal:
\begin{itemize}[nosep]
    \item Traces hot gas in galaxy clusters ($>10^6$ K)
    \item Is sensitive to AGN/supernova feedback
    \item Requires modeling both density ($n_e$) and temperature ($T$) fields
    \item Has stronger small-scale fluctuations due to gas physics
    \item Is critical for CMB experiments (Planck, ACT, SPT)
\end{itemize}

Lagrangian Deep Learning applies learned displacement fields to dark matter particles to predict the electron pressure distribution ($n_e \times T$). The model must capture:
\begin{itemize}[nosep]
    \item Gas cooling and heating processes
    \item Shock heating in dense regions
    \item AGN/stellar feedback effects
    \item Temperature-dependent ionization
\end{itemize}

\textbf{Target Field:} $n_e \times T$ [cm$^{-3}$ K] -- Electron pressure field

\textbf{Core Objective (Absolute Priority):}
\begin{itemize}[nosep]
    \item \textbf{MAXIMIZE cross-correlation $r(k)$ averaged over ALL scales} -- THIS IS THE FITNESS METRIC
    \item Maintain \textbf{numerical stability} across different configurations
    \item Ensure \textbf{physical consistency} (no negative pressure, reasonable temperature range)
\end{itemize}

\textbf{Problem Context \& Constants:}
\begin{lstlisting}[basicstyle=\ttfamily\scriptsize, frame=single]
BoxSize = cosmology['BoxSize']  # 25 Mpc/h for CAMELS
NMESH = 64                      # grid resolution
DOWNSAMPLE = 16                 # particle downsampling factor
Nstep = 3                       # number of displacement layers
TARGET = 'nT'                   # tSZ signal: electron pressure
\end{lstlisting}

\textbf{Constraints and Best Practices:}

MUST MAINTAIN:
\begin{enumerate}[nosep]
    \item Function signatures of \texttt{Displacement} and \texttt{LDL}
    \item vmad compatibility (all operations must be differentiable)
    \item MPI compatibility (distributed arrays)
    \item Numerical stability (use $\epsilon=10^{-8}$ to avoid division by zero)
    \item The EVOLVE-BLOCK-START and EVOLVE-BLOCK-END markers
\end{enumerate}

\textbf{Evaluation Metrics:}

Primary metric: \texttt{combined\_score} (higher is better)
\begin{lstlisting}[basicstyle=\ttfamily\scriptsize, frame=single]
combined_score = mean_r_k_all_scales +    # Cross-correlation r(k) averaged
                 -failed_runs * 1000.0    # Failure penalty
\end{lstlisting}

Where \texttt{mean\_r\_k\_all\_scales} is computed on:
\begin{itemize}[nosep]
    \item \textbf{Full volume} (entire simulation box, not just validation region)
    \item \textbf{All $k$ scales} (all valid $k$ modes from $k_{\min}=0.01$ to $k_{\max}=10.0$)
    \item Range: $[0, 1]$, where 1.0 = perfect correlation
\end{itemize}

\textbf{VMAD API Reference (Critical):}

vmad is NOT standard NumPy! It is a computation graph framework for automatic differentiation, similar to TensorFlow 1.x or Theano.

\textbf{Parameter Extraction -- How to get values from `param':}

\texttt{param} is a SYMBOL object, not a NumPy array. You CANNOT use normal indexing!
\begin{lstlisting}[basicstyle=\ttfamily\scriptsize, frame=single]
# CORRECT:
mu = linalg.take(param, 5*Nstep, axis=0)      # Extract single parameter
b1 = linalg.take(param, 5*Nstep+1, axis=0)    # Extract next parameter

# WRONG:
mu = param[5*Nstep]           # Error: 'Symbol' object is not subscriptable
params = param[5:10]          # Error: slicing not supported
\end{lstlisting}

\textbf{Available Operations:}

\texttt{vmad.lib.unary} -- Element-wise operations:
\begin{itemize}[nosep]
    \item Available: \texttt{exp}, \texttt{log}, \texttt{log10}, \texttt{sin}, \texttt{cos}, \texttt{tan}, \texttt{sinh}, \texttt{cosh}, \texttt{arcsin}, \texttt{arccos}, \texttt{arctan}, \texttt{absolute}, \texttt{fabs}, \texttt{sinc}
    \item NOT Available: \texttt{sqrt} (use \texttt{x**0.5}), \texttt{tanh} (use \texttt{sinh(x)/cosh(x)}), \texttt{clip}, \texttt{maximum}, \texttt{minimum}
\end{itemize}

\texttt{vmad.lib.linalg} -- Linear algebra:
\begin{itemize}[nosep]
    \item Available: \texttt{take}, \texttt{stack}, \texttt{sum}, \texttt{sumat}, \texttt{broadcast\_to}, \texttt{reshape}, \texttt{transpose}, \texttt{concatenate}, \texttt{einsum}, \texttt{matmul}, \texttt{add}, \texttt{mul}, \texttt{div}, \texttt{mod}, \texttt{pow}
    \item NOT Available: \texttt{slice}, \texttt{split}, \texttt{chunk}, \texttt{gather}, \texttt{scatter}
\end{itemize}

\texttt{vmad.lib.fastpm} -- Particle-mesh operations:
\begin{itemize}[nosep]
    \item Available: \texttt{paint}, \texttt{readout}, \texttt{decompose}, \texttt{exchange}, \texttt{gather}, \texttt{r2c}, \texttt{c2r}, \texttt{apply\_transfer}, \texttt{fourier\_space\_neg\_gradient}
\end{itemize}

\textbf{Fourier Space Operations -- Special Care Needed:}

When working in Fourier space, you must understand Symbol vs Scalar distinction:
\begin{lstlisting}[basicstyle=\ttfamily\scriptsize, frame=single]
# CORRECT PATTERN:
# Step 1: Create filter with PURE NUMPY operations in lambda
Filter = Literal(pm.create(type='complex', value=1).apply(
    lambda k, v: k.normp(2, zeromode=1e-8) ** 0.5  # Pure numpy only!
))

# Step 2: Broadcast parameters to Symbol's shape
kh = mpi.allbcast(kh, comm=pm.comm)
kh = linalg.broadcast_to(kh, vmad_eval(Filter, lambda x: x.shape))

# Step 3: Apply vmad operations on Symbols
Filter = - unary.exp(-Filter**2 / kl**2) * unary.exp(-kh**2 / Filter**2)

# WRONG PATTERN (mixing Symbol and scalar in lambda):
Filter = pm.create(type='complex').apply(
    lambda k, v: unary.exp(-(k.normp(2) / k_split)**2)  # ERROR!
)
# TypeError: must be real number, not Symbol
\end{lstlisting}

Inside \texttt{pm.apply(lambda k, v: ...)}, operations must be pure numpy. You cannot use vmad operations (\texttt{unary.*}, \texttt{linalg.*}) there!

\textbf{Common Patterns:}
\begin{lstlisting}[basicstyle=\ttfamily\scriptsize, frame=single]
# Accessing parameters:
alpha = linalg.take(param, 5*i, axis=0)
alpha = mpi.allbcast(alpha, comm=pm.comm)
alpha = linalg.broadcast_to(alpha, vmad_eval(S, lambda x: x.shape))

# Implementing missing functions:
y = x ** 0.5              # Square root via power
y = unary.sinh(x) / unary.cosh(x)  # Tanh
y = ReLU(x)               # Already defined as @operator
\end{lstlisting}

\textbf{Debugging Checklist:}
\begin{itemize}[nosep]
    \item ``AttributeError: module has no attribute'' $\rightarrow$ Function doesn't exist (e.g., \texttt{tanh}, \texttt{sqrt}, \texttt{clip})
    \item ``Symbol object is not subscriptable'' $\rightarrow$ Use \texttt{linalg.take()} instead of indexing
    \item ``Symbol has no attribute'' $\rightarrow$ Use vmad equivalents instead of numpy methods
    \item ``TypeError: must be real number, not Symbol'' $\rightarrow$ Mixed Symbol and scalar in \texttt{pm.apply()}
    \item ``operands could not be broadcast'' $\rightarrow$ Use \texttt{linalg.broadcast\_to()} after \texttt{mpi.allbcast()}
\end{itemize}
\end{tcolorbox}

The LDL codebase is built on \texttt{vmad}\footnote{\url{https://github.com/rainwoodman/vmad}}, a computation graph framework for automatic differentiation used in the FastPM ecosystem. Unlike standard NumPy, \texttt{vmad} operates on symbolic computation graphs (similar to TensorFlow~1.x or Theano), where array variables are \texttt{Symbol} objects that do not support direct indexing or in-place mutation. Because the LDL training loop requires end-to-end differentiable operations for gradient-based parameter optimization, all evolved code must be expressed using the \texttt{vmad} operator set. This non-standard API proved to be a major source of compilation failures (reflected in the low 54.5\% success rate for tSZ in Table~\ref{tab:global_statistics}), motivating the detailed API reference included in the prompt above.

\section{Using MadEvolve on New Tasks}
\label{app:new_tasks}

MadEvolve is designed to be extensible to new scientific computing tasks beyond the cosmological applications demonstrated in this paper. Detailed instructions and additional examples are available at \url{https://madevolve.org}. Applying MadEvolve to a new domain requires creating a project folder with the following components:

\textbf{1. Seed Program} (\texttt{initial.py}). The initial algorithm implementation that serves as the starting point for evolution. The code region to be evolved is marked with \texttt{\# EVOLVE-BLOCK-START} and \texttt{\# EVOLVE-BLOCK-END} comments. The LLM will propose modifications only within this block while preserving the surrounding infrastructure code.

\textbf{2. Evaluation Script} (\texttt{evaluate.py}). A script that executes the generated program on test data and outputs a \texttt{metrics.json} file containing quantitative performance measures. The evaluation should be deterministic and computationally tractable---typically completing in minutes to enable rapid iteration through generations. The primary metric used for fitness ranking must be clearly defined (e.g., $r(k)$ for BAO, $r_{\text{2D}}$ for tidal reconstruction).

\textbf{3. Evolution Configuration} (\texttt{run\_evo.py}). A Python script that configures and launches the evolution using the \texttt{EvolutionOrchestrator}. Key configuration elements include:
\begin{itemize}[nosep]
    \item \texttt{task\_description}: A detailed system prompt encoding the physical context, optimization objectives, allowed tools, and domain-specific constraints that guide the LLM's code generation
    \item \texttt{evaluator\_script}: Path to the evaluation script
    \item \texttt{init\_program\_path}: Path to the seed program
    \item Evolution hyperparameters: number of generations, population size, selection strategy, mutation modes, and parallelization settings
\end{itemize}

\textbf{4. Report Adapter} (optional). To generate automated analysis reports for a new task, one can define a scenario adapter that extends the MadEvolve analyzer framework. Each adapter implements three sub-components: (i) a \texttt{MetricsAdapter} that parses domain-specific metrics from \texttt{metrics.json} and formats comparison tables; (ii) a \texttt{PromptAdapter} that generates LLM prompts for algorithm analysis, improvement comparison, and executive summaries; and (iii) a \texttt{ReportTemplateAdapter} that defines the report structure and assembles the final document. The adapters in \texttt{MadEvolve-Cosmo/analyzer/adapters/} provide reference implementations for BAO, tidal, and LDL scenarios.\footnote{The domain-agnostic MadEvolve framework is available at \url{https://github.com/tianyi-stack/MadEvolve}. The cosmology-specific applications, prompts, evaluators, and report adapters described in this paper are at \url{https://github.com/tianyi-stack/MadEvolve-Cosmo}.}

The \texttt{examples/} directory in our repository contains complete reference implementations for BAO reconstruction, tidal field reconstruction, and LDL models that can serve as templates for new applications.

\section{LLM-generated Evolution Reports for all Tasks}\label{app:llm_report}

MadEvolve automatically generates comprehensive analysis reports for each evolutionary run, produced by the LLM-based report generator using Gemini 3 Pro Preview. The complete reports for all experiments in this paper (BAO reconstruction, 21\,cm tidal field reconstruction, LDL tSZ prediction, and the iterative BAO evolution) are available at \url{madevolve.org}.

Each report follows a standardized structure designed to provide both high-level summaries and detailed technical analysis:

\begin{enumerate}[nosep]
    \item \textbf{Executive Summary.} An LLM-generated overview of the experiment's key findings, including the magnitude of performance improvement, the nature of algorithmic innovations discovered, and implications for the relevant scientific domain.

    \item \textbf{Evolution Overview.} Quantitative statistics of the evolutionary run: total duration, number of generations and programs evaluated, success rate, the generation at which the best solution was found, and a table tracking the best fitness score across generations.

    \item \textbf{Task-Specific Quality Metrics.} Side-by-side comparison of baseline and best-evolved algorithms across all evaluation metrics, including per-scale or per-simulation breakdowns where applicable. For reconstruction tasks, this includes cross-correlation coefficients at individual wavenumber bins; for LDL, this includes per-simulation validation losses.

    \item \textbf{Baseline Algorithm Analysis.} LLM-generated scientific interpretation of the seed algorithm, covering its core approach, key physical features, implementation details, and known strengths and limitations.

    \item \textbf{Best Evolved Algorithm Analysis.} Analogous analysis of the top-performing evolved algorithm, with emphasis on novel elements that distinguish it from the baseline and a discussion of its strengths and limitations.

    \item \textbf{Evolution Improvements.} A detailed comparison identifying key modifications introduced by evolution, their physical justification, novel techniques discovered, trade-offs (e.g., computation time vs.\ accuracy, complexity vs.\ maintainability), and an assessment of scientific validity and generalization potential.

    \item \textbf{Experiment Configuration.} The hyperparameters used for the evolutionary run, including the number of islands, migration interval, maximum generations, and which LLM models were employed.

    \item \textbf{Appendix: Algorithm Code.} Full source code of both the baseline seed program and the best evolved algorithm, enabling direct inspection and reproducibility.
\end{enumerate}

\noindent Where applicable, reports additionally include tables of tunable parameters with their optimized values, bounds, and optimization method (e.g., automatic differentiation). The 21\,cm tidal reconstruction report also highlights a particularly noteworthy intermediate solution discovered at generation~52, which achieves near-optimal performance with only 9~tunable parameters compared to the 64~parameters of the overall best solution. These reports can serve as a starting point for human researchers to assess whether discovered algorithms merit further investigation and to understand the physical intuition behind the improvements.

\end{document}